\documentclass[useAMS,usenatbib]{mnras}

\usepackage{amssymb}
\usepackage{graphicx}
\usepackage{multirow}
\usepackage{longtable}
\usepackage{deluxetable}

\title[\emph{Gaia} benchmarks]{Ultracool dwarf benchmarks with \emph{Gaia} primaries}
\author[F. Marocco et al.]{F. Marocco$^{1}$\thanks{E-mail: f.marocco@herts.ac.uk}, D. J. Pinfield$^{1}$, N. J. Cook$^{1,2}$, M. R. Zapatero Osorio$^{3}$, D. Montes$^{4}$, \newauthor J. A. Caballero$^{3,5}$, M. C. G\'alvez-Ortiz$^{3}$, M. Gromadzki$^{6}$, H. R. A. Jones$^{1}$, \newauthor R. Kurtev$^{7,8}$, R. L. Smart$^{9,1}$,  Z. Zhang$^{10,11}$, A. L. Cabrera Lavers$^{10,12}$, \newauthor D. Garc\'ia \'Alvarez$^{10,12}$, Z. X. Qi$^{13}$, M. J. Rickard$^{1}$, L. Dover$^{1}$ \\ \\
$^{1}$Centre for Astrophysics Research, School of Physics, Astronomy and Mathematics, University of Hertfordshire, College Lane, \\
Hatfield AL10 9AB, UK\\
$^{2}$Faculty of Science, York University, 4700 Keele Street, Toronto, ON M3J 1P3, Canada \\
$^{3}$Centro de Astrobiolog\'ia (CSIC-INTA), Carretera de Ajalvir km 4, 28850 Torrej\'on de Ardoz, Madrid, Spain\\
$^{4}$Dpto. Astrof\'isica, Facultad de CC. F\'isicas, Universidad Complutense de Madrid, E-28040 Madrid, Spain\\
$^{5}$Landessternwarte, Zentrum f\"ur Astronomie der Universit\"at Heidelberg, K\"onigstuhl 12, 69117 Heidelberg, Germany\\
$^{6}$Warsaw University Astronomical Observatory, Al. Ujazdowskie 4, PL-00-478, Warszawa, Poland\\
$^{7}$Instituto de F\'isica y Astronom\'ia, Universidad de Valpara\'iso, Av. Gran Breta\~na 1111, Playa Ancha, Casilla 5030, Valpara\'iso, Chile\\
$^{8}$Millennium Institute of Astrophysics, Santiago, Chile\\
$^{9}$Istituto Nazionale di Astrofisica, Osservatorio Astronomico di Torino, Strada Osservatorio 20, 10025 Pino Torinese, Italy\\
$^{10}$Instituto de Astrof\'isica de Canarias, E-38205 La Laguna, Tenerife, Spain\\
$^{11}$Universidad de La Laguna, Dept. Astrof\'isica, E-38206 La Laguna, Tenerife, Spain\\
$^{12}$GTC Project Office, 38205 La Laguna, Tenerife, Spain\\
$^{13}$Shanghai Astronomical Observatory, Chinese Academy of Sciences, Shanghai 200030, China\\ 
}

\begin{document}

\date{Accepted 2017 June 14. Received 2017 June 14; in original form 2017 January 20}

\pagerange{\pageref{firstpage}--\pageref{lastpage}} \pubyear{2017}

\maketitle

\label{firstpage}

\begin{abstract}
	We explore the potential of \emph{Gaia} for the field of benchmark ultracool/brown dwarf companions, and present the results of an initial search for metal-rich/metal-poor systems. A simulated population of resolved ultracool dwarf companions to \emph{Gaia} primary stars is generated and assessed. Of order $\sim$24,000 companions should be identifiable outside of the Galactic plane ($|b| > 10\,$deg) with large-scale ground- and space-based surveys including late M, L, T, and Y types. Our simulated companion parameter space covers $0.02 \le M/M_{\odot} \le 0.1$, $0.1 \le {\rm age/Gyr} \le 14$, and $-2.5 \le {\rm [Fe/H]} \le 0.5$, with systems required to have a false alarm probability $<10^{-4}$, based on projected separation and expected constraints on common-distance, common-proper motion, and/or common-radial velocity. Within this bulk population we identify smaller target subsets of rarer systems whose collective properties still span the full parameter space of the population, as well as systems containing primary stars that are good age calibrators. Our simulation analysis leads to a series of recommendations for candidate selection and observational follow-up that could identify $\sim$500 diverse \emph{Gaia} benchmarks. As a test of the veracity of our methodology and simulations, our initial search uses UKIDSS and SDSS to select secondaries, with the parameters of primaries taken from Tycho-2, RAVE, LAMOST and TGAS. We identify and follow-up 13 new benchmarks. These include M8-L2 companions, with metallicity constraints ranging in quality, but robust in the range $-0.39 \le {\rm [Fe/H]} \le +0.36$, and with projected physical separation in the range $0.6\,<\,s/{\rm kau}\,<76$. Going forward, \emph{Gaia} offers a very high yield of benchmark systems, from which diverse sub-samples may be able to calibrate a range of foundational ultracool/sub-stellar theory and observation.
\end{abstract}

\begin{keywords}
	binaries: visual -- brown dwarfs -- stars: late type
\end{keywords}

\section{Introduction}

	Ultracool dwarfs are a mixture of sub-stellar objects that do not burn hydrogen, and the lowest mass hydrogen fusing stars. While most hydrogen-burning ultracool dwarfs (hereafter UCDs) stabilize on the stellar main-sequence after approximately 1~Gyr, their sub-stellar counterparts continuously cool down (since they lack an internal source of energy) and evolve towards later spectral types. Their atmospheric parameters are a strong function of age. The degeneracy between mass and age in the UCD regime does not affect higher mass objects \citep{1997ApJ...491..856B}.

 	Measuring directly the dynamical mass of a celestial body is possible only if the object is part of a multiple system, or via micro-lensing events. But so far the census of UCDs with measured dynamical masses is very limited \citep[see e.g.][]{2010ApJ...711.1087K,2014ApJ...790..133D,2015ApJ...805...56D}. Similarly, age indicators are poorly calibrated and, therefore, scarcely reliable, especially for typical field-star ages ($>\,1$~Gyr). 

 	The spectra of UCDs are characterized by strong alkali absorption lines, as well as by broad molecular absorption bands \citep[primarily due to water, hydrides, and methane; see e.g.][]{2005ARA&A..43..195K}. A number of these features have been shown to be sensitive to metallicity and surface gravity (both proxies for age), but the majority of studies have been so far purely qualitative \citep[e.g.][]{2001MNRAS.326..695L,2010A&A...519A..93B,2010ApJS..190..100K}, and the quantitative attempts to calibrate these age indicators suffer from large scatter and limited sample size \citep[e.g.][]{2009AJ....137.3345C,2013ApJ...772...79A} or simply do not extend all the way down through the full UCD regime \citep[e.g.][]{2007ApJ...669.1235L,2017MNRAS.464.3040Z}. Moreover, the cooling tracks for sub-stellar objects are sensitive to the chemical composition of the photosphere, further complicating the scenario \citep{1997ApJ...491..856B}. The metallicity influences the total opacity by quenching/enhancing the formation of complex molecules and dust grains, all believed to be key factors in shaping the observed spectra of sub-stellar objects. Although a number of absorption features are known to be sensitive to the total metallicity \citep[e.g.][]{2010ApJS..190..100K,2012MNRAS.422.1922P}, no robust calibration has so far been developed to determine the abundances of sub-stellar objects.

 	A way to achieve more accurate, precise and robust calibrations is to study large samples of benchmark UCD objects for which properties such as mass, age and composition may be determined/constrained in independent ways. Benchmark systems come in a variety of forms \citep[e.g.][]{2006MNRAS.368.1281P}, but here we focus on UCDs as wide companions. Such benchmark UCDs (hereafter ``benchmarks'') may be easily studied, are expected to be found over a wide range of composition and age (i.e. comparable to wide stellar binary populations), and are sufficiently common to offer large sample sizes out to reasonable distance in the Galactic disc \citep[see][]{2013MNRAS.431.2745G}. In general, system age constraints and chemical composition can be inferred from the main-sequence primaries (assuming the most likely scenario that the components formed together). This constrains the atmospheric properties of the UCD companions allowing calibration of their spectroscopic atmospheric parameter indicators. While a benchmark population has previously been found and characterized \citep[see e.g.][]{2011EPJWC..1606012D,2014ApJ...792..119D,2015ApJ...802...37B,2015MNRAS.454.4476S,2016ApJS..224...36K,2017MNRAS.466.2983G}, their number remains limited and the parameter space is therefore largely under-sampled. 

 	The advent of the European Space Agency (ESA) cornerstone mission \emph{Gaia} \citep{2016A&A...595A...1G} provides the potential to greatly expand the scope/scale of benchmark studies. Combined with the capabilities of deep wide-field infrared surveys optimizing sensitivity to distant UCD companions, \emph{Gaia} will yield exquisite parallax distances and system property constraints \citep[e.g.][]{2003AAS...202.5505B} for an unprecedented sample of benchmark systems. Indeed, to take full advantage of the \emph{Gaia} benchmark population within a reasonable programme of follow-up study, we aim to identify a subset with a focus on covering the full range of UCD properties (i.e. biased towards outlier properties). This sample should reveal the nature of UCDs extending into rare parameter space, i.e. high and low metallicity, youthful and ancient, and the coolest UCDs.

	To access this outlier benchmark population it is crucial to identify systems in a very large volume. Wide companions can be confirmed through an assessment of their false alarm probability, using a variety of parameters. System components should have an approximate common distance, since the orbital separation is much less than the system distance, as well as common proper motion and approximately common radial velocity, since the orbital motion should be small compared to system motion. Previous studies have focused on common distance and proper motion, but across the full \emph{Gaia} benchmark population we may use a different compliment of parameters. In particular, for more distant systems proper motion will be smaller and radial velocity may be more useful.

 	Once discovered, benchmark systems need to be characterized via detailed spectroscopic studies of both the primary stars and their sub-stellar companions. While \emph{Gaia} will provide (in addition to astrometry) radial velocity and atmospheric parameter estimates for the primaries, most UCDs will be too faint to be detected by the ESA satellite. Even those UCDs that are bright enough to be astrometrically observed by \emph{Gaia} will be too faint for its Radial Velocity Spectrometer. So further study of the UCDs will be needed to determine proper motions, radial velocities, spectral indices, and metallicity/age indicators necessary to fully exploit these benchmarks.

 	In this paper we explore the full scope of the expected \emph{Gaia} benchmark population, and then present discoveries from our first selection within a portion of the potential parameter space. Sections~\ref{sims} and \ref{sim_res_disc} describe a simulation we performed of the local Galactic disk, containing wide UCD companions to \emph{Gaia} stars as well as a population of field UCDs. We simulate constraints on benchmark candidates (using appropriate limits set by \emph{Gaia} and available deep large-scale infrared surveys), and calculate false alarm probabilities (based on a range of expected follow-up measurements) thus identifying the full benchmark yield within our simulation. We assess the properties of this population, address a series of pertinent questions, and determine how best to optimize a complete/efficient identification of the full population of benchmark systems for which \emph{Gaia} information will be available. In Sections~\ref{cand_sel}, \ref{spectroscopy}, \ref{proper_motion}, and \ref{new_systems} we then outline and present discoveries from our initial search. We have targeted systems where the \emph{Gaia} primary has metallicity constraints (from the literature), and where the UCD companion is a late M or L dwarf (detected by the United Kingdom Infrared Deep Sky Survey, hereafter UKIDSS, or the Sloan Digital Sky Survey, hereafter SDSS) with an on-sky separation $\le\,3$~arcmin from its primary. Conclusions and future work are discussed in Section~\ref{discussion}.

\section{Simulation}
	\label{sims}
 	Our simulation consists of both a field population of UCDs and a population of wide UCD companions to \emph{Gaia} primary stars. This two-component population allowed us to simulate the calculation of false alarm probabilities (the likelihood of field UCDs mimicking wide companions by occupying the same observable parameter space), and thus identify simulated benchmark UCDs that we would expect to be able to robustly confirm through a programme of follow-up study.

	\subsection{The field population \label{field_pop}}
		We simulated the UCD field population within a maximum distance of 1~kpc, and over the mass range $0.001 < M/M_\odot < 0.12$ (a parameter space that fully encompasses our detectable population of UCD benchmarks; see Section~\ref{section_mass_age_spt} and \ref{section_dist}). The overall source density is normalized to 0.0024~pc$^{-3}$ in the 0.1$-$0.09~$M_\odot$ mass range, following \citet{2008A&A...486..283D}, and consistent with the values tabulated by \citet{2008A&A...488..181C}. We simulated the field population across the whole sky with the exception of low Galactic latitudes (i.e. $|b| < 10\,$deg), since detecting UCDs in the Galactic plane is challenging due to high reddening and confusion \citep[see e.g.][]{2012MNRAS.427.3280F,2017MNRAS.464.1247K}. 

		Each UCD is assigned a mass and an age following the \citet{2005ASSL..327...41C} log-normal Initial Mass Function (hereafter IMF) and a constant formation rate. In \citet{2012ApJ...754...30P} it is shown that the \citet{2005ASSL..327...41C} IMF describes very well the $\sigma$ Orionis observed mass function, except for the very low mass domain ($M\,\lesssim\,0.01\,M_\odot$), where the discrepancy becomes increasingly large. However the difference is at very low masses, where the number of expected detections is low given the observational constraints (see Section~\ref{sim_res_disc}). 

		The observable properties of the UCD are determined using the latest version of the BT-Settl models (\citealt{2003A&A...402..701B} isochrones in the $0.001\,<\,M\,<\,0.01\,M_\odot$ mass regime, and \citealt{2015A&A...577A..42B} isochrones in the $0.01\,<\,M\,<0.12\,M_\odot$ mass regime). The isochrones are interpolated to determine $T_{\rm eff}$, log~$g$, radius, and the absolute UKIDSS, SDSS, and \emph{Wide-field Infrared Survey Explorer} \citep[\textit{WISE},][]{2010AJ....140.1868W} magnitudes. 

		The UCDs are then placed in the Galaxy by generating a set of XYZ Cartesian heliocentric coordinates in the same directions as UVW Galactic space motions (X positive towards the Galactic centre, Y positive in the direction of Galactic rotation, and Z positive towards the north Galactic pole). We assume a homogeneous distribution in X and Y \citep[similar to previous work; e.g.][]{2006MNRAS.371.1722D}. Although the nearest spiral arm is located at $\sim$800~pc \citep[Sagitarius--Carina spiral arm, see][]{2015MNRAS.450.4150C}, the most distant of our simulated benchmark population are actually at $\sim$550~pc (see Section~\ref{section_dist}), and our assumption should thus be reasonable. The distribution in Z follows the density laws adopted by the \emph{Gaia} Universe Model Snapshot \citep[GUMS, see Table 2 in][]{2012A&A...543A.100R}. The XYZ coordinates are then converted to right ascension ($\alpha$), declination ($\delta$), and distance using standard transformations. 

		We assigned to each UCD the $UVW$ components of its velocity by drawing them from a Gaussian distribution centered on zero. The velocity dispersions ($\sigma_U , \sigma_V ,$ and $\sigma_W$ respectively) depend on the age of the UCD and are taken, for consistency, from \citet[Table 7]{2012A&A...543A.100R}. $V$ was corrected for the asymmetric drift, also following \citet{2012A&A...543A.100R}. $UVW$ are then converted to proper motion and radial velocity using standard transformations. 

		Apparent magnitudes for our simulated objects were calculated by applying the distance modulus ignoring reddening and extinction, which should be low-level since our simulated objects are not at low Galactic latitude and are within the local volume. We included unresolved binaries within our sample by assuming a 30$\%$ binary fraction \citep[e.g.][]{2015MNRAS.449.3651M} and that all unresolved binaries are equal-mass \citep[a reasonable approximation according to e.g.][]{2007prpl.conf..427B}.

		Limiting this field simulation to $J\,<\,19$~mag or $W2\,<\,15.95$~mag (i.e. the same photometric limits we apply to our simulated benchmarks; Section~\ref{sim_constraints}), produces a population of $\sim\,1,700,000$ UCDs.

	\subsection{The \emph{Gaia} benchmarks population \label{bench_pop}} 
		We generated a model population of benchmark systems by selecting random field stars, and adding one UCD companion around a fraction of them. Primary stars were chosen randomly from GUMS, assuming the fraction of L dwarf companions to main sequence stars, in the $30-10,000\,$au separation range, to be 0.33$\%$, as measured by \citet{2013MNRAS.431.2745G}. Note that \citet{2013MNRAS.431.2745G} only measured the fraction of main sequence stars hosting L dwarf companions, and here our simulation assumes the same system-fraction for initially injected companions around all types of primaries (i.e. main sequence stars, white dwarfs, giants and sub-giants). The fraction of stars hosting late-M, T, and Y dwarfs follows from the above normalisation coupled with our other simulated characteristics. More details on the simulation of benchmark systems are given in the following sub-sections. 

		\subsubsection{Primaries: \emph{GUMS} \label{gums}} 
			The primary stars of our benchmark systems are selected from GUMS \citep{2012A&A...543A.100R}. The detailed description of GUMS can be found in \citet{2012A&A...543A.100R}, and here we only briefly summarize the relevant facts. GUMS represents a snapshot of what \emph{Gaia} should be able to see at an arbitrary given epoch. As such, it contains main-sequence stars, giants and sub-giants, white dwarfs, as well as rare objects (Be stars, chemically peculiar stars, Wolf-Rayet stars, etc.), thus providing us with a diverse and reasonably complete sample of potential primaries. The stars were generated from a model based on the Besan\c con Galaxy model \citep[hereafter BGM;][]{2003A&A...409..523R}. Since the BGM produces only single stars, binaries and multiple systems were added in with a probability that increases with the mass of the primary star, and orbital properties following the prescriptions of \citet{2011AIPC.1346..107A}, resulting in a fraction of binary systems within 10~pc of $24.4\,\pm\,0.4\%$ \citep{2011AIPC.1346..107A}. Exoplanets are added around dwarf stars, following the probabilities given by \citet{2005ApJ...622.1102F} and \citet{2009ApJ...697..544S}, and with mass and period distributions from \citet{2002MNRAS.335..151T}. GUMS does not include brown dwarfs. It is important to note here that GUMS generated stars in several age bins, following a constant formation rate over the $0-10\,$Gyr range, with the addition of three bursts of star formation at 10, 11, and 14~Gyr representing the Bulge, Thick disk, and Spheroid respectively.

			We selected GUMS primaries within 1~kpc of the Sun and with $|b| > 10\,$deg, and down to $G\,<\,20.7\,$mag \citep[i.e. the limit for \emph{Gaia} detection;][]{2016A&A...595A...2G,2017arXiv170309454S}. No other restrictions are placed on the selected objects. The full sample of GUMS potential primaries amounts to $\sim\,63,000,000$ stars.

		\subsubsection{Companions \label{companions}} 
			For our randomly assigned companion population, we assigned masses following the \citet{2005ASSL..327...41C} IMF. Distance, age, metallicity, proper motion and radial velocity values were set using the associated primary stars. Similarly to our simulated field population, $T_{\rm eff}$, log~$g$, radius, and the absolute MKO, SDSS, and \textit{WISE} magnitudes were calculated for the companions using the BT-Settl models \citep{2003A&A...402..701B,2015A&A...577A..42B}. 

			For our main simulation the frequency of UCD companions was assumed to be flat with the logarithm of projected separation ($s$). This is reasonably consistent with observations over the range where completeness is high \citep[$<\,10\,$kau; see e.g.][]{2014ApJ...792..119D}. Companions are assigned out to $s\,=\,50$~kau, with limited constraints by previous observations on the wider part of this range \citep[see e.g.][]{2009A&A...507..251C}. However, we account for the truncation of wide companions through dynamical interaction (see below), which provides a more physical means of shaping the frequency distribution of the widest benchmark companions. And note that we also carried out a ``re-run'' simulation with the frequency of UCD companions declining linearly with the log of projected separation, more closely matching observations across the full range (see Section~\ref{priority} and Table~\ref{sim_res_summary}), although it is our main simulation that we discuss in detail in Section~\ref{priority}. The position of the UCD companion in the sky, relatively to the primary, is generated assuming a homogeneously distributed position angle. 

			Dynamical interactions between stars are known to cause the disintegration of multiple systems \citep{1987ApJ...312..367W}. This is particularly critical in the case of our simulated wide benchmarks. The chance of a system undergoing such disintegration increases as a function of time, and can be estimated, given the system total mass and age, using the method of \citet{2010AJ....139.2566D}. The average lifetime $\tau$ of a binary system is given by

			\begin{equation}
			\tau \simeq 1.212\, \frac{M_{\rm tot}}{a}\,{\rm Gyr}
			\label{time}
			\end{equation}

			\noindent where $M_{\rm tot}$ is the total mass of the binary in units of $M_\odot$ and $a$ is the semi-major axis in pc. We removed from our simulated sample all systems whose age is greater than their expected lifetime. To convert from $s$ to $a$ we assumed a randomly distributed inclination angle between the true semi-major axis and the simulated projected separation. While this is obviously an approximation (a complete treatment would take into account the full set of orbital parameters), it leads to a median $a/s\,=\,1.411$, closely matching the 1.40 ratio derived from theoretical considerations by \citet{1960JO.....43...41C}. 

			Particular care was taken while treating UCD companions to white dwarfs (WDs). In that case the total mass of the system and separation change as the main sequence progenitor evolves into a WD. Therefore we first estimated the cooling age of the WD using its $T_{\rm eff}$ and log~$g$ (given by GUMS) and the cooling tracks for DA WDs from \citet{2011ApJ...730..128T} \citep[since all WDs in GUMS are assumed to be DAs,][]{2012A&A...543A.100R}. We estimated the mass of the WD progenitor using the initial-to-final mass relation of \citet{2008MNRAS.387.1693C}. We assumed that the orbit of a companion around a star that becomes a WD, will expand stably, such that

			\begin{equation}
			\frac{a_{\rm WD}}{M_{\rm WD}}\,=\,\frac{a_{\rm MS}}{M_{\rm MS}}
			\end{equation}

			\noindent where $M_{\rm WD}$ and $M_{\rm MS}$ are the mass of the WD and of its progenitor, and $a_{\rm WD}$ and $a_{\rm MS}$ are the semi-major axes of the orbit in the WD and main sequence phase, respectively. We then calculate a ``disintegration probability'' for each stage as follows

			\begin{equation}
				p_{\rm WD}\,=\,\tau_{\rm cool}/\tau_{\rm WD}
			\end{equation}

			\begin{equation}
				p_{\rm MS}\,=\,\tau_{\rm star}/\tau_{\rm MS}
			\end{equation}

			\noindent where $\tau_{\rm cool}$ is the cooling age of the WD, $\tau_{\rm star}$ is the ``main sequence age'' (the difference between the total age given by GUMS and $\tau_{\rm cool}$), $\tau_{\rm WD}$ is the expected lifetime in the ``WD-stage'' (i.e. assuming $a\,=\,a_{\rm WD}$ and $M_{\rm tot}\,=\,M_{\rm WD}\,+\,M_{\rm UCD}$ in equation \ref{time}), and $\tau_{\rm MS}$ is the expected lifetime in the ``main sequence stage'' (i.e. assuming $a\,=\,a_{\rm MS}$ and $M_{\rm tot}\,=\,M_{\rm MS}\,+\,M_{\rm UCD}$ in equation \ref{time}). We removed systems whose total disintegration probability $p_{\rm WD}\,+\,p_{\rm MS}$ is greater than one.

			We note that the initial-to-final mass relation only holds in the mass range $0.5\,<\,M_{\rm WD}/M_\odot\,< 1.1$. For WDs more massive than 1.1~$M_\odot$, we simply assume the main sequence lifetime of the progenitor to be negligible compared to the cooling age of the WD, since $M_{\rm MS}$ would be greater than 6~$M_\odot$. For WDs less massive than 0.5~$M_\odot$, we assume the mass loss during the post-main-sequence evolution to be negligible, hence $M_{\rm MS} \simeq M_{\rm WD}$. 

 	\subsection{Simulating constraints on candidate selection and follow-up} 
 	\label{sim_constraints}
 		After generating the field and benchmark populations, we simulated limitations on UCD detection within infrared surveys, as well as the accuracy of observational follow-up. This involves imposing magnitude and minimum separation cuts, and generating realistic uncertainties (typically achieved) on the observables (magnitudes, distance, RV and proper motion). 

 		The first step is to set detection limits for our simulated UCDs. Current near- and mid-infrared (hereafter NIR and MIR) surveys probe the sky at different depths and with different levels of multi-band coverage, but rather than try to simulate all these different surveys (which would be convoluted, and may change in the future) we took a somewhat simplified approach. In the near-infrared we chose a depth limit of $J\,\leq\,19$~mag, which can be achieved in a variety of ways. The ongoing Visible and Infrared Survey Telescope for Astronomy (hereafter VISTA) Hemisphere Survey \citep[VHS,][]{2013Msngr.154...35M} is scanning the Southern hemisphere down to $J\,=\,21.2$~mag and $K_s\,=\,20.0$~mag, allowing for the detection and selection of UCD candidates, e.g. via $J\,-\,K_s$ colour criteria. In the northern hemisphere, the combination of the UKIDSS Large Area Survey (with limiting magnitude $J\,=\,19.5$~mag), SDSS, UKIDSS Hemisphere Survey (with $J$ depth similar to UKIDSS LAS) and Pan-STARRS~1 \citep{2016arXiv161205560C} will allow the effective selection of UCD candidates (using e.g. $z\,-\,J$ criteria) across the full hemisphere.

 		While NIR surveys should be ideal to select most UCDs, some with very red near-mid infrared colours (particularly Y dwarfs) will be best detected in the MIR. \textit{WISE} is scanning the whole sky down to a $5\,\sigma$ limit of $W2\,=\,15.95$~mag. We can therefore expect to identify UCDs down to this limit, by selecting $W2$-only detections or objects with very red $W1-W2$ colours. These objects would be much fainter in the NIR bands, however the spectroscopic follow-up of \emph{WISE}-selected targets (down to $J\,\sim\,21-22\,$mag) is routinely achieved with the aid of the latest generation of $6-8\,$m-class telescopes \citep[e.g.][]{2011ApJ...743...50C,2012ApJ...753..156K,2013ApJ...776..128K,2014MNRAS.444.1931P}. 

 		Any simulated object (either in the field or part of a benchmark system) fainter than $J\,=\,19.0$~mag and $W2\,=\,15.95$~mag is therefore considered undetectable and removed from our simulated population. 

 		Since we are only targeting resolved star+UCD systems, we need to remove all unresolved companions. The angular resolution of existing NIR and MIR surveys is rather patchy, varying from $1-2\,$arcsec in the best cases (e.g. SDSS, VISTA, UKIDSS) to $\sim 6$~arcsec for \emph{WISE}. Additionally, large area surveys are known to have issues identifying and cataloging sources around bright stars, pushing the detection limit for faint companions out to larger separations. We chose to adopt an ``avoidance radius'' dependent on brightness, i.e. an area of sky around a star were faint UCDs will go undetected. Examination of a range of example stars in SDSS (where this effect is quite clear) led us to set the following values: 

 		\begin{itemize}
 			\item 15 arcmin for stars with $V\,\leq\,4$~mag 
 			\item 10 arcmin for stars with $4\,<\,V\,\leq\,6$~mag
 			\item 5 arcmin for stars with $6\,<\,V\,\leq\,8$~mag
 			\item 4 arcsec for stars with $V\,>\,8$~mag (if $J\,\leq\,19$~mag)
 			\item 20 arcsec for stars with $V\,>\,8$~mag (if $J\,>\,19$~mag and W2~$\leq\,15.95$~mag)
 		\end{itemize}

 		Simulating realistic uncertainties on distance, proper motion and radial velocity is a complicated exercise, since it depends not only on the brightness of the UCD, but also on the type of follow-up assumed. For instance, dedicated astrometric campaigns can achieve a high level of precision on parallax and proper motion down to very faint magnitudes \citep[$\sim\,1$~mas down to $J\,\sim\,20\,$mag; e.g.][]{2013Sci...341.1492D,2013MNRAS.433.2054S}, but are time consuming and limited to a relatively small number of objects. However, we take a simplified approach since it is more common to measure proper motion for UCDs using just two epochs, i.e. following up the original discovery images at a later epoch allowing a long enough time baseline. With second epoch images often obtained with a different telescope/filter, the precision of such measurements is limited. With medium-to-high resolution spectroscopy one can obtain radial velocities down to a precision of a few km~s$^{-1}$ or less \citep[e.g.][]{2007ApJ...666.1205Z,2010ApJ...723..684B} but these observations are limited to the brighter objects only. Following these considerations, we adopted a 10~mas~yr$^{-1}$ uncertainty for proper motions (assuming a two-epoch measurement and a $\sim 5$~yr baseline), and a precision of 2~km~s$^{-1}$ for radial velocities down to $J\,=\,18$~mag \citep{2015MNRAS.449.3651M}. For fainter objects we consider a radial velocity measurement to be currently unfeasible, and we therefore assume their radial velocity to be unconstrained. 

 		To simulate distance uncertainties, we considered the current spectrophotometric distance calibrations. Although based on an increasing number of UCDs with measured parallaxes \citep[see e.g.][]{2010A&A...524A..38M,2012ApJS..201...19D}, these calibrations are limited by the intrinsic scatter in the UCD population, primarily due to age and composition differences among objects of similar spectral type. The typical scatter around the polynomial spectrophotometric distance relations is $\sim\,0.4$~mag \citep[][]{2012ApJS..201...19D}. We thus adopted distance modulus uncertainties of 0.4~mag. No systematic uncertainty is considered for e.g. young or peculiar objects. For unresolved binaries (in our simulated field population), where a spectrophotometric relation would lead to an incorrect distance estimate, we assume their observed distance to be 30\% closer than their real distance, with distance modulus uncertainties of 0.4~mag.

	\subsection{Companionship probabilities and ``confirmable \emph{Gaia} benchmarks'' \label{prob}}

		We determined a series of companionship probabilities for each simulated benchmark system, appropriate for the observable properties that would be available at each stage of a search-and-follow-up programme. This companion confirmation programme was represented through the following stages: (i) cross matching the GUMS primary with the simulated field + benchmark UCDs out to the separation of the simulated companion, to account for both cross-contamination (i.e. a UCD companion to star A being erroneously associated with nearby star B), and potential companion mimics whose spectrophotometric distance is consistent with the parallax distance of the primary (within $2\,\sigma$); (ii) obtaining the proper motions of the candidate primary and companion and ensuring they are consistent (within $2\sigma$); (iii) obtaining the radial velocities of the candidate primary and companion and ensuring these too are consistent (within $2\sigma$). 

		While the $2\sigma$ distance criteria is prone to contamination from background unresolved binary UCDs (whose underestimated spectrophotometric distance may fall within the $2\sigma$ distance range of the primary), it also makes it unlikely that unresolved binary UCD companions will be mistakenly rejected. This is because such unresolved binaries are overluminous by no more than 0.75~mag (the equal-mass limit) which is within two times our adopted $\sigma\,=\,0.4$~mag uncertainty.

		For each benchmark UCD companion, ``mimics'' were sought in the field population (i.e. UCDs that meet the observational requirements for companionship). This simulate-and-search exercise was carried out 10,000 times following a Monte-Carlo approach, for each search-and-follow-up stage. A ``false alarm probability'' was then determined equal to the number of trials where at least one mimic was found divided by 10,000. And the companionship probability was set as one minus the false alarm probability. Our approach cannot accurately calculate very small ($<$0.01\% but non-zero) false alarm probabilities, however, it is effective at identifying systems with a strong companionship probability. We chose a minimum threshold for companionship probability of 99.99\% (close to 4-$\sigma$ confidence) for the confirmation of simulated benchmarks. Some benchmark systems were confirmed after early stages of our search-and-follow-up (see discussion in the next section), but we made our full selection of confirmed simulated systems by applying the threshold at the final stage. We refer to this full sample as ``confirmable \emph{Gaia} benchmarks''.

\section{Simulation results and discussion} 
	\label{sim_res_disc}
 	We now discuss the results of our simulated population of confirmable \emph{Gaia} benchmarks (CGBs). Primarily we consider our main simulation (resulting from a projected separation distribution that is flat in log~$s$ out to $50$~kau), which likely represents an upper bound on the overall population size. However, at the end of this discussion we present CGB subset sizes for both flat and sloping separation distributions, where the sloping distribution is a closer match to observations (albeit with biases and selection effects), and thus provides a likely lower bound.

	The output of our main simulation consists of 36,559 ultracool companions (with $T_{\rm eff}\,<\,2800$~K) with $J\,\leq\,19$~mag or $W2\,\leq\,15.95$~mag. When we consider our statistical requirements to confirm companionship this reduces to 24,196 CGBs. In the following subsections we discuss the distribution of this CGB sample within intrinsic and observable parameter space, and then consider prioritized CGB subsets and an optimized search-and-follow-up approach.

	\subsection{Intrinsic properties of confirmable \emph{Gaia} benchmarks \label{intr_prop}}

		\begin{figure}
			\includegraphics[width=0.5\textwidth]{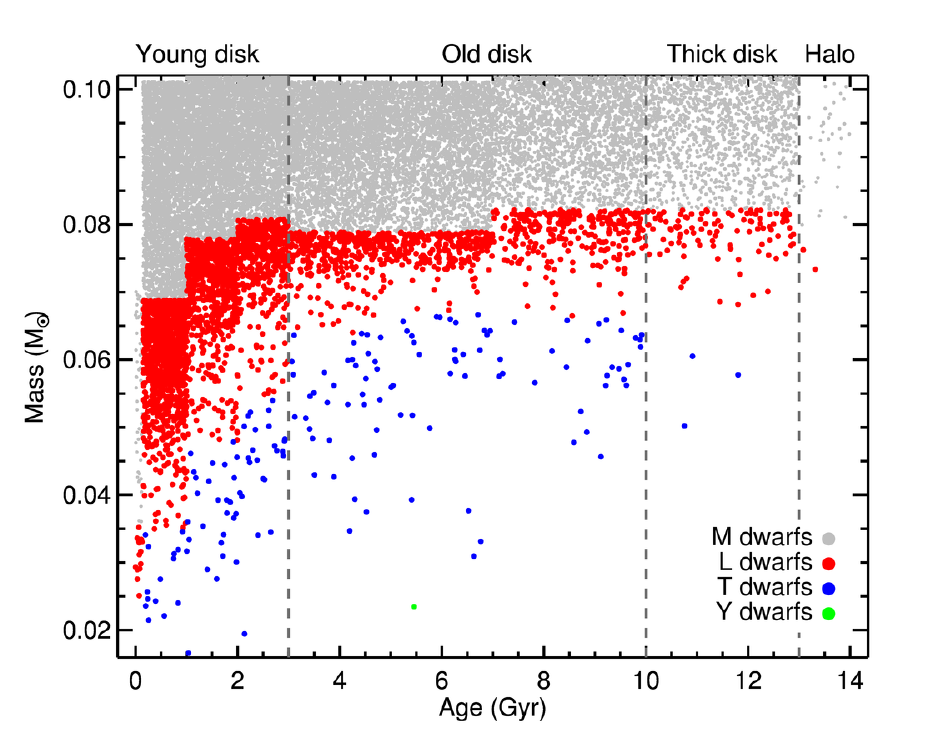}
			\caption{The mass-age distribution of our simulated CGBs, with spectral types M7--M9, L, and T plotted in grey, red and blue respectively \label{mass_age_mlt}}
		\end{figure}

		\subsubsection{Mass, age, and spectral type \label{section_mass_age_spt}} 
			Figure~\ref{mass_age_mlt} shows the mass-age distribution of the CGBs. For each of the GUMS age bins we have introduced a random scatter (across the bin) so as to obtain a more natural continuum of ages and make the plot easier to view. Note that this leads to ``step-like'' behaviour as one moves across the age bins. The M-L and L-T spectral type transitions can be seen separating the grey-red and red-blue plotting colours. The large majority of CGBs are low-mass stars, with 82\% having masses above the sub-stellar limit and 18\% being brown dwarfs. The lowest mass CGBs are found in the youngest age bin with masses down to $\sim\,0.02\,M_\odot$. Most CGBs (87\%) are ultracool M types, with about 13\% having L or T spectral type. The predominance of late M type CGBs is due to a combination of three factors: (i) M dwarfs are brighter and therefore can be seen out to larger distance given our adopted magnitude limits; (ii) M dwarfs are intrinsically more numerous given the adopted IMF; (iii) M dwarf companions are generally more massive than L and T dwarfs, and are therefore more likely to survive dynamical disruption (a less significant factor, but not negligible).

			At the oldest extremes the large number of M type CGBs includes 37 halo systems (with ages $>\,13$~Gyr). Of the 2,987 L type CGBs, about two-thirds are young disk, and $\sim\,30\,\%$ old disk. There are also 110 thick-disk L type CGBs, but a very limited number of halo L types (just 2 simulated CGBs). 

			Most of the 160 T type CGBs are nearly evenly split between the young and old disk populations (43\% and 52\% respectively), with a small but potentially interesting collection of 3 T type CGBs in the thick-disk. Our simulation does not predict any T type CGBs in the halo. At young ages there are 67 late M and L-type objects $<\,500$~Myr, but no T-type objects in this age range. Our most youthful age bin ($<\,100$~Myr) contains 52 M types and 15 L dwarfs. 

			Our main simulation does contain one Y dwarf (with $T_{\rm eff}\,=\,490$~K). Although WD~0806-661~B \citep{2011ApJ...730L...9L} is a known wide Y dwarf companion to a white dwarf (discovered in \textit{Spitzer} data), it lies beyond our all-sky photometric limits. Our one simulated Y CGB would be within the \textit{WISE} All-Sky survey, and would be bright enough for spectroscopic follow-up with current facilities (Section~\ref{priority} provides further discussion on the potential for Y dwarf CGBs).

			These results are summarized in Table~\ref{sim_res_summary}, and overall suggest a potentially very large CGB population. Although dominated by low-mass stars and late M dwarfs, there should be numerically substantial samples of L and T CGBs across a wide range of age and kinematic population.

			\begin{figure}
				\includegraphics[width=0.5\textwidth]{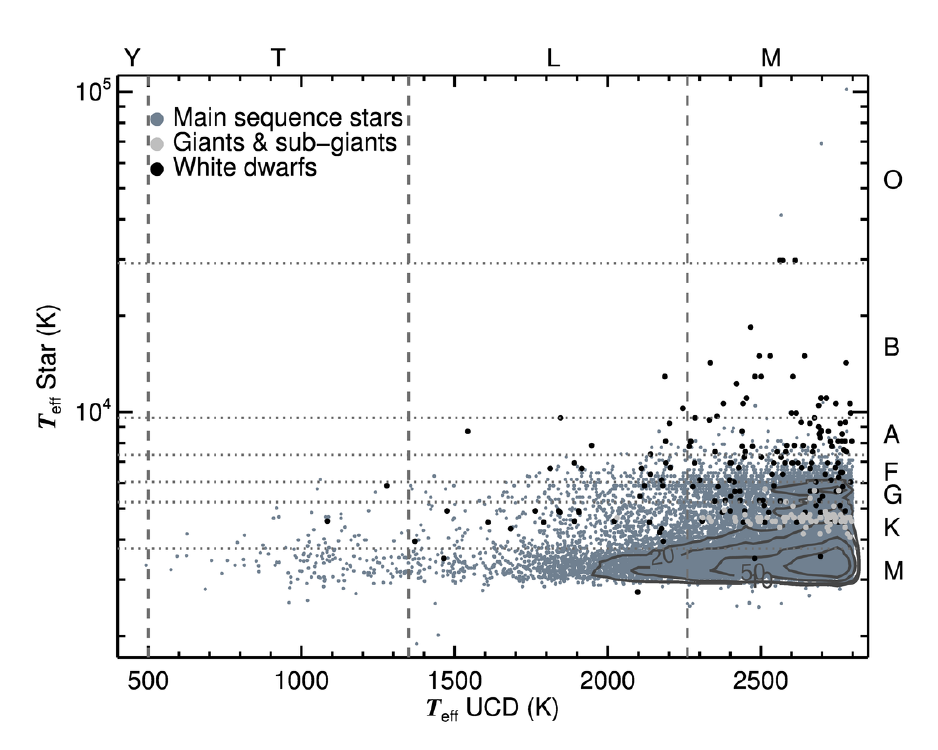}
				\caption{The distribution of secondary versus primary $T_{\rm eff}$ for our simulated CGB population. Spectral type divisions are indicated along the top and right axis. Different primary types are plotted in different shades (dark grey, light grey and black for main-sequence, sub-giant and white dwarfs respectively). We overplot density contours over the main sequence samples to show the structure of the distribution in highly crowded regions. Contour labels are, from the outermost to the innermost, 10, 20, 50, and 100 objects per $25\,\times\,100\,{\rm K}^2$ bin. \label{teff_teff}}
			\end{figure}

		\subsubsection{Binary constituents \label{bin_constituents}} 
			Figure~\ref{teff_teff} shows the distribution of secondary versus primary $T_{\rm eff}$ for the CGB population. About 60\% of CGBs have M dwarf primaries, with primary $T_{\rm eff}$ down to $\sim\,3000$~K ($\sim$~M5). Most of the remainder have FGK primaries, with just 76 CGBs containing hotter BA type primaries. In addition to the main-sequence primaries there are 75 CGBs with sub-giant primaries, and 172 with white dwarf primaries. Below $T_{\rm eff}\,\sim\,3000$~K we observe a sharp drop in the number of primaries. This is because the absolute magnitude sequence for M dwarfs is very steep for optical bands \citep[dropping nearly 3~mag between M5 and M7 in the SDSS $r$ band;][]{2011AJ....141...98B}, and therefore the $G\,<\,20.7$~mag limit results in a sharp cutoff in the population. As was discussed by \citet{2006MNRAS.368.1281P}, sub-giants and white dwarfs make very useful benchmark primaries. It is possible to constrain the metallicity and ages of sub-giant stars quite accurately (as they evolve relatively quickly across the HR-diagram) using well understood models. White dwarf primaries provide lower-limit system ages from their cooling age. Furthermore, higher mass white dwarfs have higher mass shorter lived progenitors and the cooling ages will be a better proxy for total system age.

			\begin{figure}
				\includegraphics[width=0.5\textwidth]{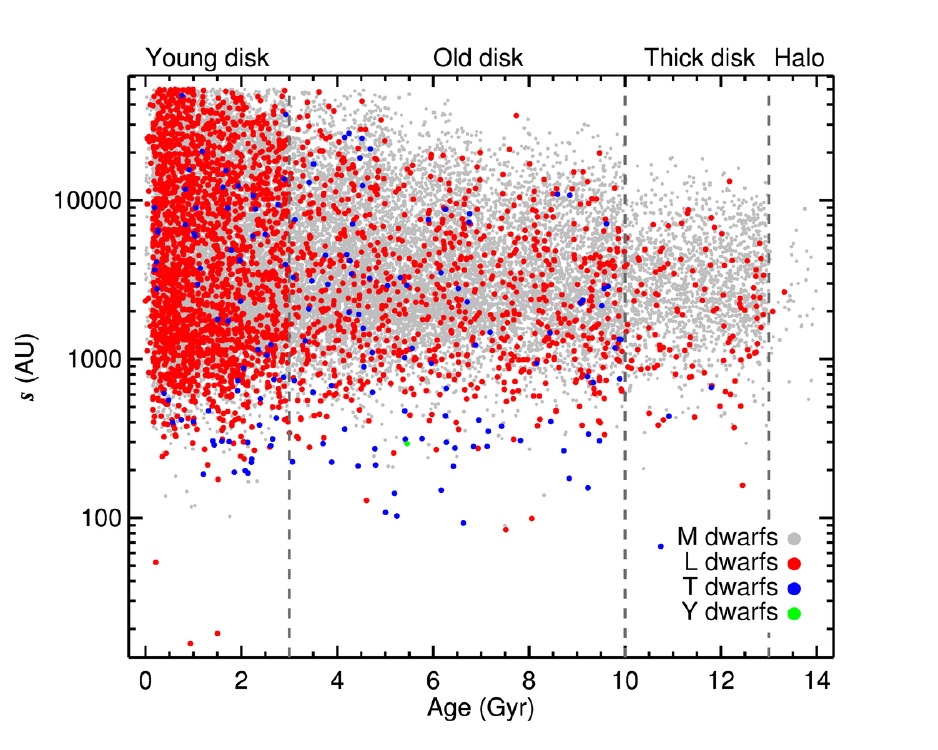}
				\caption{The projected separation $s$ versus age distribution of our simulated CGBs. UCD spectral types are coloured as in Figure~\ref{mass_age_mlt}\label{age_phys_sep}}
			\end{figure}

		\subsubsection{Projected separation} 
			Figure~\ref{age_phys_sep} shows the projected separation versus age distribution of the CGBs, with UCD spectral types coloured as in Figure~\ref{mass_age_mlt}. As we described in Section~\ref{companions} our initial separation distribution is flat in log~$s$, and is truncated at 50~kau. This truncation is seen in Figure~\ref{age_phys_sep}, as is the effect of dynamical break-up which removes CGBs if their age exceeds the (mass and separation sensitive) dynamical-interaction lifetime. This dynamical effect essentially leads to a reduced truncation across the old-disk, thick-disk and halo, but also thins the CGB population for separations greater than a few thousand au. It is interesting to note that the dynamical interaction lifetime limits all thick disk CGBs to separations $<\,20$~kau, and all halo CGBs to separations $<\,10$~kau. While our input assumptions about the separation distribution have some inherent uncertainties, our simulation results provide some useful constraints on suitable limits for the separation of CGBs across a range of kinematic populations.

			\begin{figure}
				\includegraphics[width=0.5\textwidth]{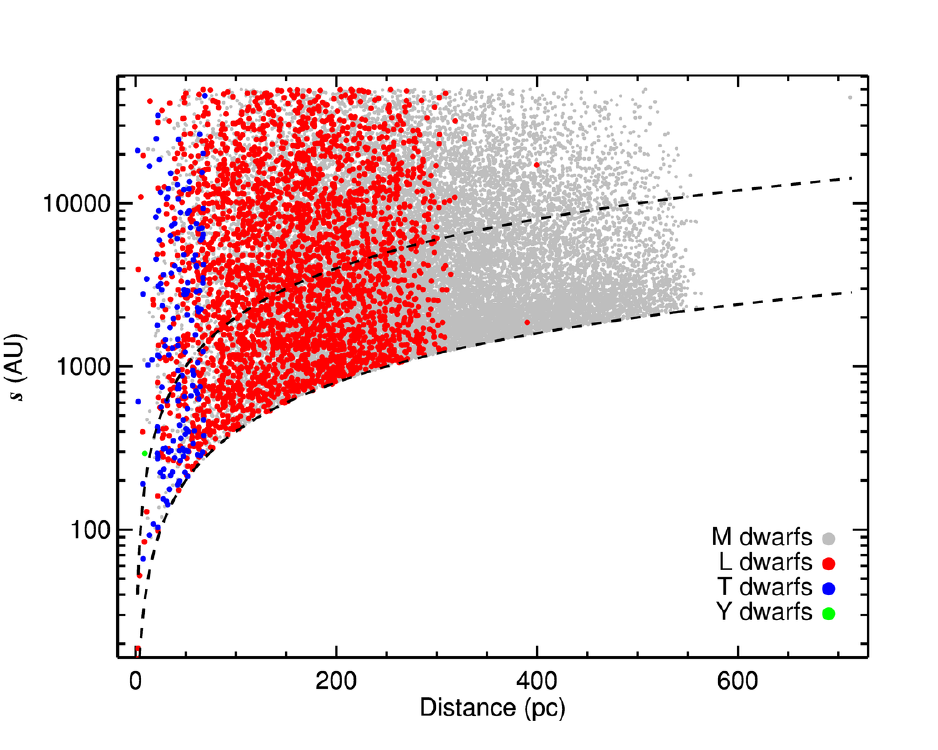}
				\caption{The projected separation versus distance distribution of our simulated CGBs. UCD spectral types are coloured as in Figure~\ref{mass_age_mlt}. The two dotted lines represent the 4~arcsec and 20~arcsec limits (see section~\ref{sim_constraints}). \label{dist_phys_sep}}
			\end{figure}

		\subsubsection{Distance \label{section_dist}} 
			Figure~\ref{dist_phys_sep} shows the distance versus projected separation distribution of the CGBs, with UCD spectral types coloured as in Figure~\ref{mass_age_mlt}. Late M, L, and T type CGBs are available out to distances of $\sim\,550$~pc, 400~pc and 70~pc respectively, and with numbers very limited for distances $<$~20~pc. There is a fairly uniform increase in the number of L type CGBs over the distance range 50--250~pc, since the increase in space volume at larger distance is counteracted by the decrease in the range of L sub-types that are detectable at this distance (i.e. all L CGBs can be detected at $\sim$~50~pc whereas only early L CGBs are detectable at $\sim$~400~pc; see also \citealt{2008A&A...488..181C}). Dashed lines delineate regions where CGBs are undetectable in the near-infrared and mid-infrared surveys because they are unresolved from their primaries (as dictated by our minimum angular separations limits of 4 and 20~arcsec respectively; see Section~\ref{companions}). The near-infrared angular resolution limit has a much greater impact because the majority of CGBs are detectable in the $J$-band (this will be discussed further in Section~\ref{obs_prop}).

			\begin{figure}
				\includegraphics[width=0.5\textwidth]{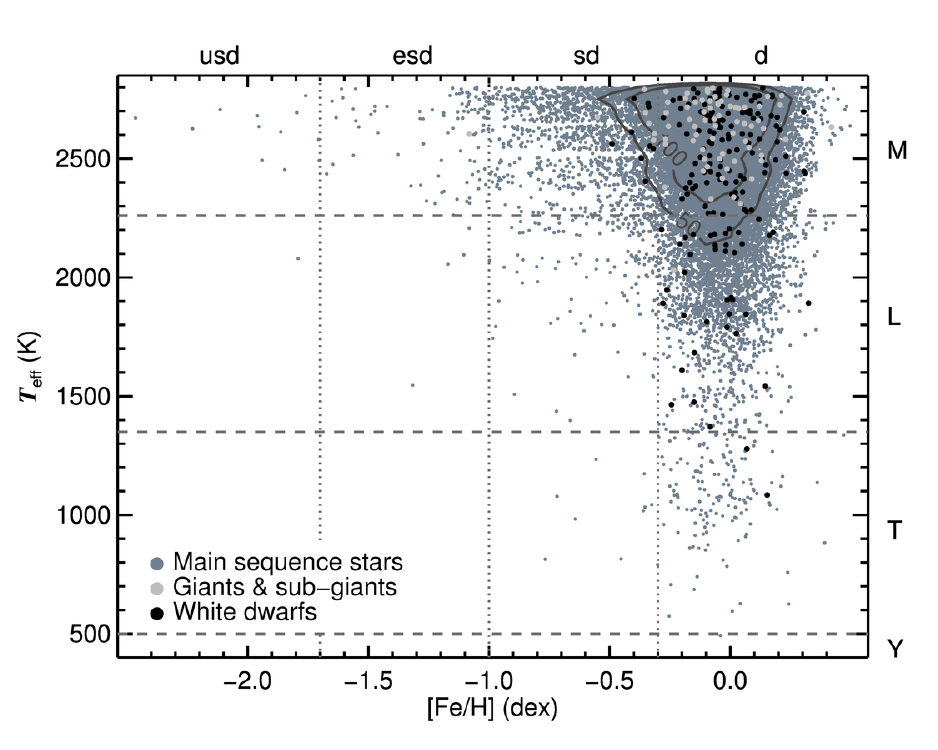}
				\caption{Effective temperature versus metallicity for our simulated CGBs. The approximate metallicity class ranges indicated along the top axis (with \textsf{d} standing for dwarf, \textsf{sd} for sub-dwarf, \textsf{esd} for extreme sub-dwarf and \textsf{usd} for ultra sub-dwarf) are based on \citet{2007ApJ...669.1235L} and \citet{2017MNRAS.464.3040Z}. Different primary types are shaded as in Figure~\ref{teff_teff}. We overplot density contours over the main sequence samples to show the structure of the distribution in highly crowded regions. Contour labels are, from the outermost to the innermost, 50 and 100 objects per $0.1$~dex~$\times\,25$~K bin. \label{teff_feh}}
			\end{figure}

		\subsubsection{Metallicity} 
			Figure~\ref{teff_feh} shows UCD $T_{\rm eff}$ versus metallicity for the CGBs. UCD spectral type divisions and approximate metallicity class ranges \citep{2007ApJ...669.1235L,2017MNRAS.464.3040Z} are indicated along the right and top axis, with sd standing for ``sub-dwarf'', esd for ``extreme sub-dwarf'', and usd for ``ultra sub-dwarf''. There is a sizable subset of 735 metal rich ([Fe/H]$>\,0.2$~dex) M type CGBs, and a smaller but significant subset of 104 metal rich L types (though there are very few metal rich T types). Within metallicity classes a large subset of 2,098 sdM CGBs should be available, with smaller subsets of 99 esdM and 12 usdM CGBs. In addition there is a subset of 149 sdL CGBs, as well as 4 esdL and 1 usdL types. Our simulation also contains 10 sdT CGBs. For these metal rich/poor CGBs about 53\% have M dwarf primaries and most of the remainder have FGK primaries (as was discussed in Section~\ref{bin_constituents}, and summarised in Table~\ref{sim_res_summary}). CGBs with sub-giant or white dwarf primaries are predominantly solar metallicity dwarfs. Most of the CGBs with sub-giant primaries are late M type (save for one L type). CGBs with white dwarf primaries are mostly ($\sim\,80\%$) late M type \citep[cf.][]{2008MNRAS.388..838D}, with the remainder generally L type, except for 2 T type CGBs \citep[cf.][]{2011MNRAS.410..705D}.

	\subsection{Observable properties of confirmable \emph{Gaia} benchmarks \label{obs_prop}}

		\subsubsection{\emph{Gaia} combined with infrared surveys}

			\begin{figure}
				\includegraphics[width=0.5\textwidth]{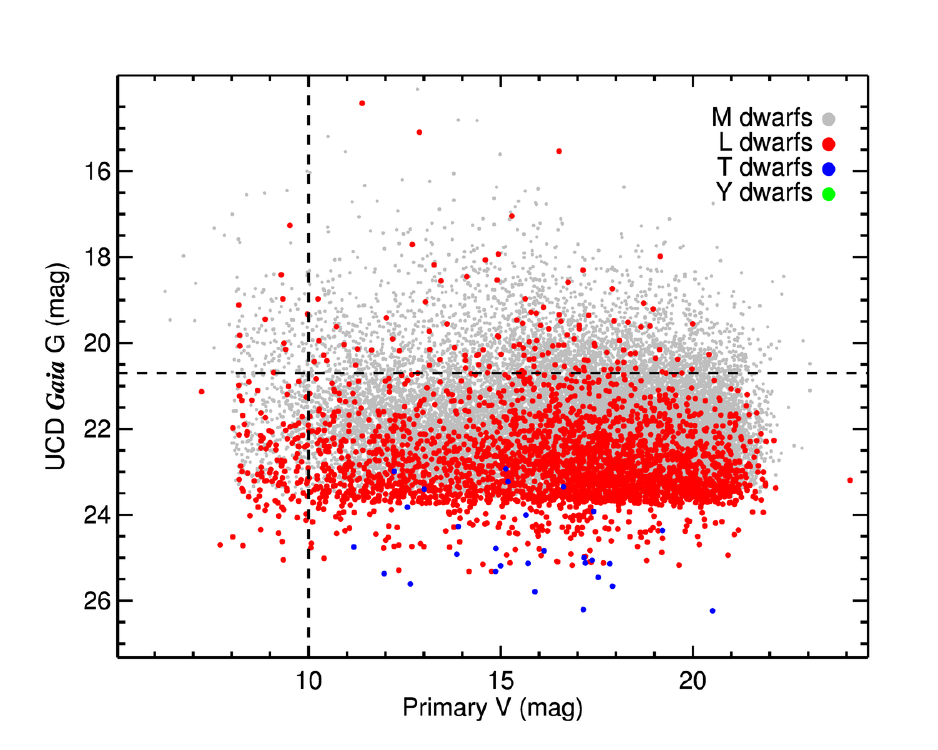}
				\caption{The \emph{Gaia G}-band magnitudes of CGBs versus $V$-band primary magnitudes, with UCD spectral types coloured as in Figure~\ref{mass_age_mlt}. Dashed lines indicate the \emph{Hipparcos} limit ($V\,\sim\,10$~mag) for the primaries, and the \emph{Gaia G} detection limit for the CGBs. \label{V_G}}
			\end{figure}

			Figure~\ref{V_G} shows the \emph{Gaia} $G$-band magnitudes of CGBs versus $V$-band primary magnitudes, with UCD spectral types coloured as in Figure~\ref{mass_age_mlt}. Dashed lines indicate the \emph{Hipparcos} limit ($V\,\sim\,10$~mag) for the primaries, and the \emph{Gaia} $G$ detection limit for the CGBs. To determine how the use of \emph{Gaia} primaries improves over samples with \emph{Hipparcos}/Gliese primaries, we counted simulated CGBs in which the primaries have $V\,<\,10$~mag or distance $<\,25$~pc. This produced 583 \emph{Hipparcos}/Gliese systems, including 129 L dwarfs and 18 T dwarfs. The entire simulated CGB sample thus represents a forty-fold increase over samples with \emph{Hipparcos}/Gliese primaries. Also, to compare the approach of using infrared surveys (for CGB detection) to detecting UCDs with \emph{Gaia} itself, we counted simulated CGBs with \emph{Gaia} $G\,<\,20.7$~mag giving 2,960 systems. Most of these are late M dwarfs, with 125 L dwarfs and no T dwarfs. The infrared surveys thus improve CGB sample-size by ten-fold for late M, and thirty-fold for L dwarfs compared to \emph{Gaia} alone. They also provide sensitivity to $>\,150$ T type CGBs that are undetectable with \emph{Gaia}.

			It is also interesting to compare the predictions of our simulation with the results of previous work to identify large samples of wide binaries using ground based surveys that are deeper than \textit{Hipparcos}/Gliese. Here we focus on the Sloan Low-mass Wide Pairs of Kinematically Equivalent Stars \citep[SLoWPoKES;][]{2010AJ....139.2566D}. SLoWPoKES uses the SDSS DR7 and requires both components of each system to be SDSS-detected. While SLoWPoKES \citep{2010AJ....139.2566D} adopted common proper motion criteria, SLoWPoKES-II \citep{2015AJ....150...57D} identified associations through common distance only. Their distances however were based on photometric calibrations only, and therefore they had to restrict their angular separation limit to a maximum of 20~arcsec in order to obtain false alarm probabilities $<\,5\%$. SLoWPoKES-II identified 43 wide companion UCDs as well as 44 wide UCD binaries (formed of two UCDs). If we compare their results with a ``SLoWPoKES-like'' sample drawn from our simulation \citep[i.e. requesting common distance, maximum separation of 20~arcsec, and imposing magnitude limits at $z\,<\,20.5$~mag and $i\,<\,21.3$~mag;][]{2015AJ....150...57D} we find 380 UCD companions, 21 of which are L dwarfs. This represents a four-fold increase over SLoWPoKES-II, presumably due to their growing incompleteness towards the faint magnitude limits. Even in the absence of such incompleteness, NIR surveys allow access to a much larger sample of late L- and T-type companions that are entirely precluded from red-optical surveys. Moreover, the use of \emph{Gaia} astrometry allows one to target UCD companions out to much wider angular separations, thanks to the improved distance constraints.

		\subsubsection{Magnitude and angular separation \label{mag_ang_sep}}

			\begin{figure}
				\includegraphics[width=0.5\textwidth]{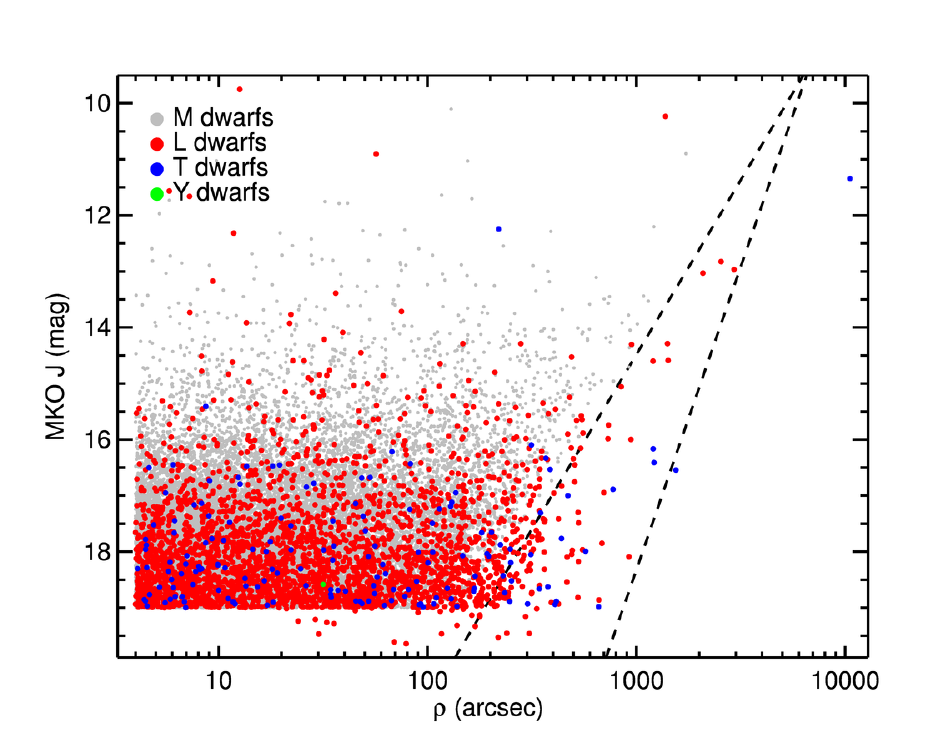}
				\caption{$J$-band magnitude versus angular separation ($\rho$) for our simulated CGBs, with UCD spectral types coloured as in Figure~\ref{mass_age_mlt}. At wide separation increased false alarm probability leads to truncation in the CGB population (see Section~\ref{mag_ang_sep}), with hard cut-offs (for the M and L systems) shown by dashed lines. \label{J_ang_sep}}
			\end{figure}

			Figure~\ref{J_ang_sep} shows $J$-band magnitude versus angular separation for the CGBs, with UCD spectral types coloured as in Figure~\ref{mass_age_mlt}. The $J\,=\,19$~mag selection cut-off (see Section~\ref{field_pop}) is the limiting factor for the vast majority of the CGBs, although there are a small number of L type with $J\,>\,19$~mag. These are brighter than our mid-infrared limit ($W2\,<\,15.95$~mag), but have $(J-W2)\,>\,3$~mag \citep[i.e some L5--T3 dwarfs, and most T7+ dwarfs; see Figure~7 from][]{2011ApJS..197...19K}. The late M and L type CGBs become far more numerous at fainter $J$ due to the increased space volume at larger distance (particularly for the earliest types). The large majority of M types have $J\,=\,16-19$~mag, and most L types have $J\,=\,17.5-19$~mag. The T type CGBs are nearly uniformly spread across the range $J\,=\,16-19$~mag. For the M and L types it can be seen that the maximum angular separation becomes more truncated towards fainter $J$-band magnitude, and that this effect is strongest for the M types (see dashed lines). This truncation is caused by the increased false-alarm-probability at larger distance and wider angular separation (since the space volume in which such CGB mimics may be found is larger). This affects the more distant M types most, and the closer T types least. Although there are a few CGBs with angular separations of a few degrees, the large majority have smaller separation, with most M and L types $<\,3-5$~arcmin, and most T types $<\,15$~arcmin.

			\begin{figure}
				\includegraphics[width=0.5\textwidth]{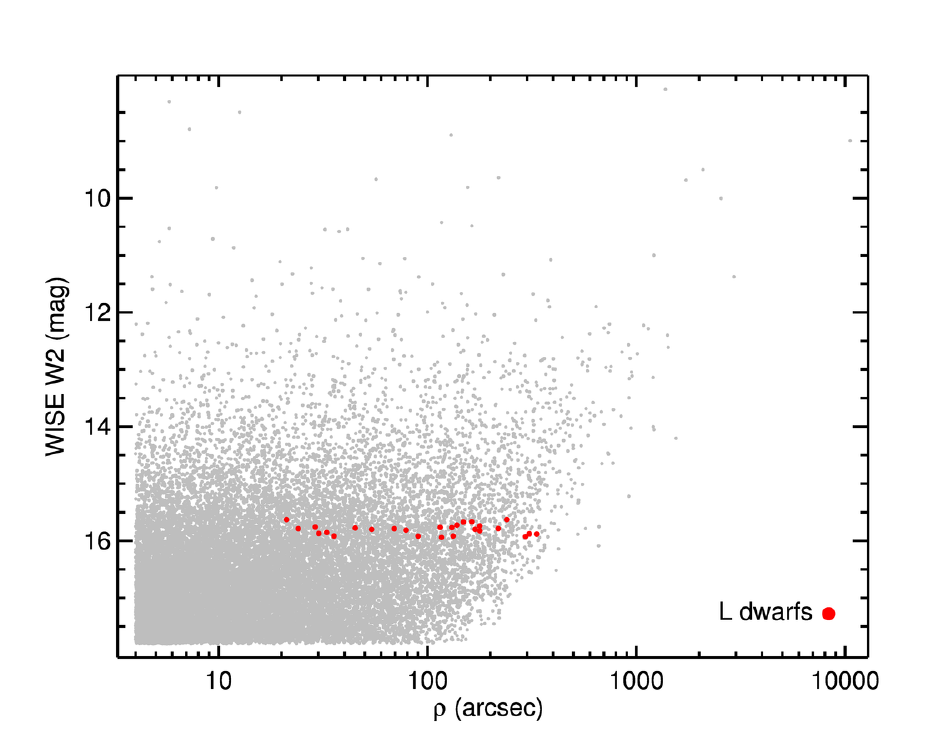}
				\caption{$W2$-band magnitude versus angular separation ($\rho$) for our simulated CGBs. L type CGBs with $J\,>\,19$~mag are coloured red (with other CGBs coloured grey). \label{W2_ang_sep}}
			\end{figure}

			Figure~\ref{W2_ang_sep} shows the $W2$-band magnitude versus angular separation, high-lighting CGBs that only pass our MIR threshold (i.e. $W2\,\leq\,15.95$~mag and $J\,>\,19$~mag). As Figure~\ref{J_ang_sep} also showed there are only a small number of L type CGBs (and no T types) with $J\,>\,19$~mag, but Figure~\ref{W2_ang_sep} makes it clear that there is a large majority of CGBs that are fainter than the $W2$ limit. The sharp cutoff at $W2\,\sim\,17.8$~mag is due to the combination of the $J\,=\,19$~mag cut, combined with the typical colours of UCDs. For $W2\,>\,17.5$~mag the population is dominated by distant late Ms, and their typical $J\,-\,W2\,\sim\,1.2$~mag results in the observed $\sim\,17.8$~mag cutoff.

		\subsubsection{Proper motion and radial velocity}

			\begin{figure*}
				\includegraphics[width=0.49\textwidth]{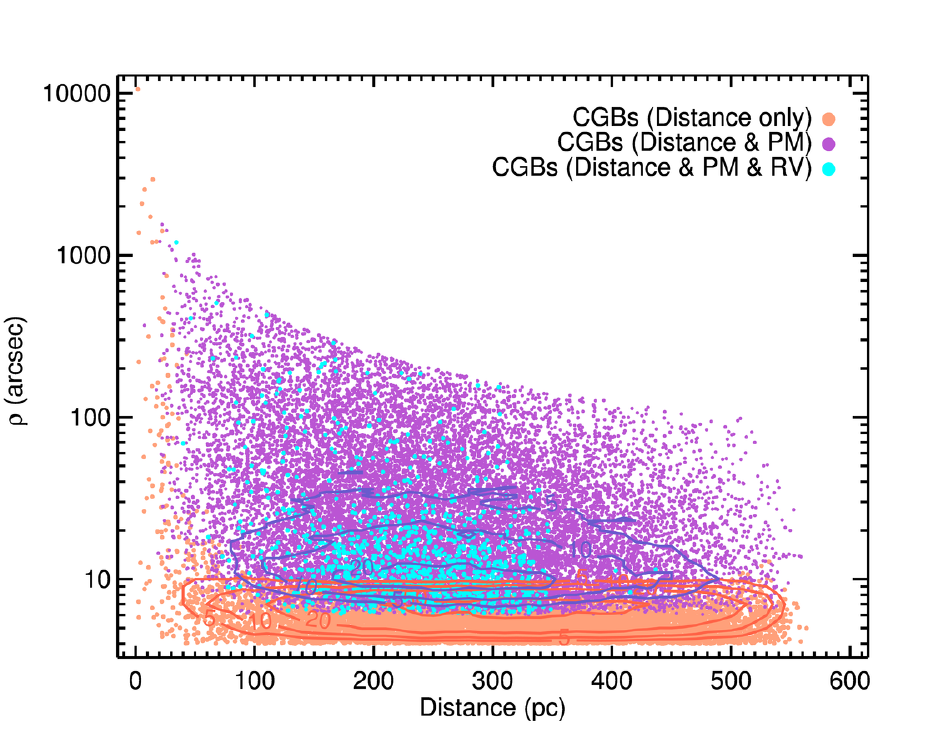}
				\includegraphics[width=0.49\textwidth]{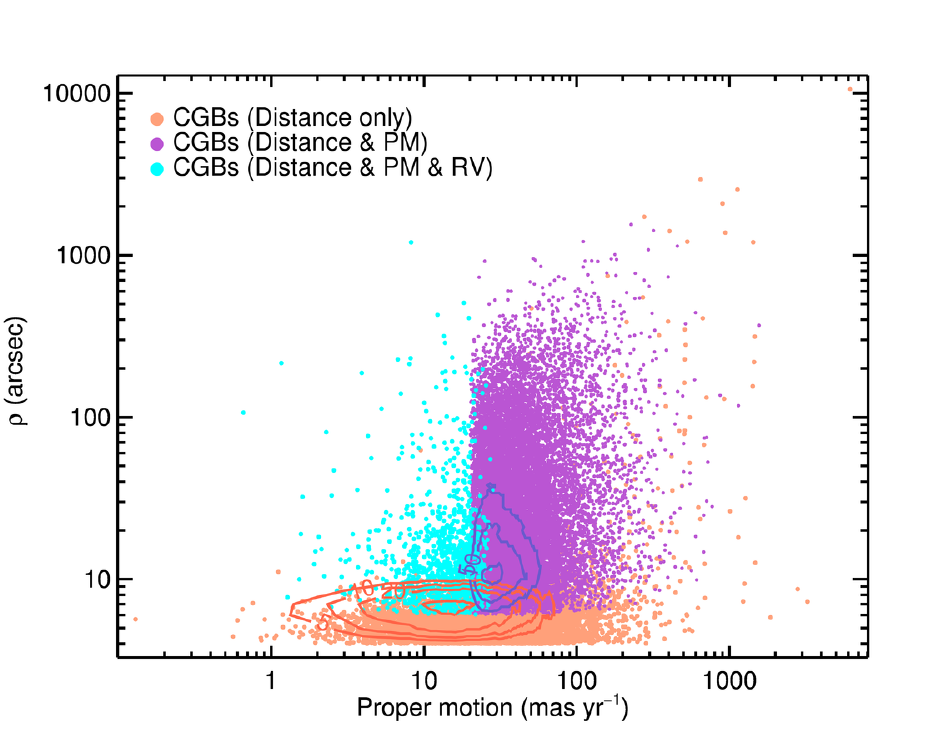}
				\caption{Angular separation ($\rho$) versus distance (left panel), and $\rho$ versus total proper motion (right panel) for the CGBs. Benchmarks that acquired CGB status (i.e. false-alarm-probability $<\,0.0001$) based only on common-distance mimics are shown in orange. Those that acquired CGB status based on common-distance and common-proper motion mimics are shown in purple. And those that also required common-RV assessment are shown in cyan. We overplot density contours over the ``distance only'' and the ``distance \& PM'' samples to show the structure of the distribution in highly crowded regions. Contour labels are 5, 10, 20, and 50 objects per $10$~pc~$\times\,1$~arcsec bin in the left panel, and objects per 1~mas~yr$^{-1}\,\times\,1$~arcsec bin in the right panel. \label{dist_pm_sep}}
			\end{figure*}

			Figure~\ref{dist_pm_sep} shows angular separation versus distance (left panel), and angular separation versus total proper motion (right panel) for the CGBs. Symbols have been colour coded to indicate those that acquired CGB status (i.e. false-alarm-probability $<\,0.0001$) through common distance criteria alone (orange), and those that also required common proper motion (purple) and common-radial velocity (cyan). For any CGB with angular separation $\rho\,<\,6$~arcsec one only needs to confirm common-distance, because even for the most distant CGBs the volume for mimics is never large enough to produce false positives. At distances $d\,<\,30$~pc, Figure~\ref{dist_pm_sep} shows that common-distance-only CGBs (orange) may be found out to wider angular separation, since the volume for mimics shrinks as distance decreases. Above 6~arcsec angular separation and 30~pc distance, it can be seen in the left panel that proper motion is generally important for confirming CGBs. And if the proper motion is small ($\mu\,<\,20-25$~mas~yr$^{-1}$), the right panel shows that RV may also need to be measured.
 
 	\subsection{Priority benchmark subsets and optimized discovery \label{priority}}

		We have shown that the CGB population is very large ($\sim$24,000 strong), covers a wide range of parameter space, and constitutes a variety of primary/secondary combinations and properties that have been discussed in Section~\ref{intr_prop} and \ref{obs_prop}. In Table~\ref{sim_res_summary} we summarise expected numbers for CGBs that could be available in a variety of subsets that are a priority for benchmark studies. These numbers are presented as ranges, encompassing our main simulation (previously discussed), and a ``re-run simulation'' in which we adopt a separation distribution that is sloping in log~$s$. This re-run simulation is a closer match to observations \citep[e.g.][]{2014ApJ...792..119D} albeit with a range of observational bias and selection effects. The slope we chose is $-11.7$, which is a rough fit to the distribution in Figure~13 from \citet{2014ApJ...792..119D}. Thus the ranges we present in Table~\ref{sim_res_summary} reflect uncertainties in the CGB population, but should encompass likely outcomes. 

		Within each subset numbers are broken down for M, L and T type. In addition, the final column in Table~\ref{sim_res_summary} indicates the expected number of CGBs in each subset that could be unresolved doubles \citep[i.e. consist of an unresolved binary UCD in a wide orbit around a primary star; e.g.][]{2014ApJ...790..133D}. Such benchmarks can yield dynamical UCD masses as well as age and compositional constraints. These numbers were determined by normalising our subset sizes using the expected fraction of unresolved binary UCDs in a magnitude limited sample. We determined this fraction to be 32\% using our simulated field population.

		As an extension to our main simulation discussion we further considered the potential for Y dwarf CGBs. Although just one was predicted in our main simulation, this was limited by the typical W2 depth of the AllWISE catalogue ($W2\,=\,15.95$~mag). This catalogue was constructed from images collected between 2010 Jan 7 and 2011 Feb 01, amounting to slightly more than two complete sky coverages. However, since \emph{WISE} was re-activated in 2013 Dec 13 it has been obtaining two additional sky coverages per year (approximately annual NEOWISE data releases), and by the end of the \emph{Gaia} mission there could be six years of additional \emph{WISE} imaging to complement the AllWISE dataset. For regions of \emph{WISE} sky around nearby stars, images could be offset and stacked (so as to be co-moving with the nearby stars), and thus provide an extra magnitude of photometric depth. 

		We have therefore considered the possibility that Y-dwarf CGBs could be identified down to a photometric depth of $W2\,\sim\,17$~mag, and note in Table~\ref{sim_res_summary} that this extension of our main simulation suggests that several ($\sim\,1-3$) Y-dwarf CGBs could be available ($T_{\rm eff}\,=\,360-480$~K). There is a rich diversity observed amongst the known Y dwarf population, an explanation for which would benefit greatly from even a small population of CGBs.

		\begin{table*}
			\begin{tabular}{l l c c}
    		Primary & Companion & Single & Doubles \\
    		 & & (min--max) & (min--max) \\
    		\hline
    		\hline
    		\multirow{3}{*}{Main sequence stars} & M dwarfs & 16,462--20,842 & 5,268--6,669 \\
    		    & L dwarfs & 2,392--2,948 & 765--943 \\
    		    & T dwarfs & 121--159 & 39--51\\
    		\hline
    		\multirow{3}{*}{Sub-giants} & M dwarfs & 76--74 & 24--24 \\
    			& L dwarfs & 5--1 & 2--0\\
    			& T dwarfs & 0 & 0 \\
    		\hline
    		\multirow{3}{*}{White Dwarfs} & M dwarfs & 96--132 & 31--42 \\
    			& L dwarfs & 25--38 & 8--12 \\
    			& T dwarfs & 0--2 & 0--1 \\
    		\hline 
    		\multirow{3}{*}{Metal-rich stars ([Fe/H]$\,>\,0.2$~dex)} & M dwarfs & 539--735 & 172--235 \\
				& L dwarfs & 84--104 & 27--34 \\
				& T dwarfs & 7--7 & 2--2 \\
    		\hline
    		\multirow{3}{*}{Metal-poor stars ([Fe/H]$\,<\,-0.3$~dex)} & M dwarfs & 1717--2209 & 549--707 \\
    			& L dwarfs & 169--154 & 54--49 \\
    			& T dwarfs & 10--10 & 3--3 \\
    		\hline
    		\multirow{3}{*}{Young stars ($<\,500$~Myrs)} & M dwarfs & 35--52 & 11--17 \\
    			& L dwarfs & 9--15 & 3--5 \\
    			& T dwarfs & 0 & 0 \\
    		\hline
    		\multirow{3}{*}{Thick disk stars} & M dwarfs & 1251--1635 & 400--523 \\
    			& L dwarfs & 111--110 & 35--35 \\
    			& T dwarfs & 8--3 & 3--1 \\
    		\hline
    		\multirow{3}{*}{Halo stars} & M dwarfs & 28--37 & 9--12 \\
    			& L dwarfs & 2--2 & 1--1 \\
    			& T dwarfs & 1--0 & 0 \\
    		\hline
    		Any & Y dwarfs & 0(1)--1(3)$^a$ & 0(0)--0(1)$^a$ \\
    		\hline
    		\hline
  			\end{tabular}
  			\caption{Properties of the Confirmable \emph{Gaia} Benchmarks (CGBs) generated by our simulation. The first column indicates the type of primary, the second column indicates the type of companion. In the third column we present the number of companions generated by our simulation when assuming a sloping separation distribution (minimum of range) and a log-flat separation distribution (maximum of range). The last column indicates the number of companions that could be expected to be unresolved binaries themselves, estimated assuming a binary fraction of 32\% (see Section~\ref{priority} for details). Notes -- $^a$ The numbers in brackets assume an extension to the main simulation reaching $W2\,\sim\,17$~mag through a shift-and-stack approach with multiple NEOWISE scans of the sky. \label{sim_res_summary}} 
		\end{table*}

		In order to identify and confirm CGBs in the priority subsets that we have high-lighted in Table~\ref{sim_res_summary}, we summarise below some recommended search-and-follow-up guide-lines that build on our discussion in Section~\ref{obs_prop}.

	\begin{itemize}
  		\item{In general, near-infrared surveys such as UKIDSS, UHS, and VISTA can yield almost all CGBs.}
  		\item{The AllWISE database adds a relatively small number of L type CGBs beyond what can be identified using near-infrared surveys.}
  		\item{The combination of AllWISE and NEOWISE imaging could improve mid-IR sensitivity and yield a sample of $\sim\,1-3$ Y-dwarf CGBs.}
  		\item{Most CGBs can be identified out to an angular separation of $3-5$~arcmin for M and L dwarfs, and $\sim\,$15~arcmin for T dwarfs.}
  		\item{Ultracool halo M dwarf CGBs all have angular separations $\rho\,<\,1$~arcmin.}
  		\item{Priority subsets can be sought in different ways. One could search for CGB candidates around target primaries (i.e. sub-giants, white dwarfs, metal rich/poor main sequence stars, young stars, thick disk and halo stars).}
  		\item{Comprehensive lists of interesting primaries may come from various sources, including \emph{Gaia} photometric and spectroscopic analysis.}
  		\item{As a complementary approach (assuming non-comprehensive primary information), one could seek observationally unusual UCDs (e.g. colour or spectroscopic outliers) that may be more likely members of our priority subsets, and search around them for primary stars.}
  		\item{A follow-up programme to confirm CGBs can be guided by Figure~\ref{dist_pm_sep}. Candidate CGBs will have known angular separation from their potential primaries, and \emph{Gaia} will provide distance constraints and total proper motions for these primaries. CGB confirmation is likely to require only common-distance (i.e. by measuring UCD spectral type) if $\rho\,<\,6$~arcsec, and in most cases when $d\,<\,30$~pc; beyond these limits common-proper-motion will be needed, provided that the proper motion is significant ($\mu\,\geq\,20\,-\,25$~mas~yr$^{-1}$); for low proper motion systems ($\mu\,<\,20\,-\,25$~mas~yr$^{-1}$) common-RV may be required.}
	\end{itemize}

\section{Candidate selection}
	\label{cand_sel}
	We now report the first results of our effort to search and follow up UCDs from the priority benchmark subsets defined in Section~\ref{priority}. Our first search was made with a bias towards metal-rich and metal poor CGBs. Prior to the release of a complete \emph{Gaia} sample we identified a list of possible primaries from several sources. We chose possible metal-rich ([Fe/H]$\,>\,0.2$~dex) and metal-poor ([Fe/H]$\,<\,-0.3$~dex) stars from a collection of catalogues. From the VizieR database\footnote{\url{http://vizier.u-strasbg.fr/}} we selected all catalogues containing distance measurements and metallicity constraints from spectroscopy or narrow band photometry (this compilation is summarised in Table~\ref{primary_summary}). We also used the 4th Data Release of the \textit{RAdial Velocity Experiment} \citep[RAVE DR4,][]{2013AJ....146..134K}, which provides spectroscopic metallicity measurements. And we selected from the compendium of photometric metallicities for 600,000 FGK stars in the Tycho-2 catalogue \citep{2006ApJ...638.1004A}, which contains estimated fundamental stellar properties for Tycho-2 stars based on fits to broadband photometry and proper motion. In addition to these metallicity-biased selections, we also included Data Release 2 of the \textit{Large sky Area Multi-Object fiber Spectroscopic Telescope} \citep[LAMOST DR2,][]{2015MNRAS.448..855Y}. To further expand our list of primaries into the M dwarf regime we included the photometric/proper motion selected catalogue of \citet[hereafter NJCM]{2016MNRAS.457.2192C}. Our possible primary list thus explores a small fraction of CGB parameter space, is significantly limited in photometric depth (by comparison to a complete \emph{Gaia} sample), and has a range of uncertainties for metallicity and distance constraints. We filtered TGAS measurements \citep[Tycho$-$\emph{Gaia} Astrometric Solution,][]{2016A&A...595A...2G,2016A&A...595A...4L} into our analysis when assessing candidate follow-up results (in Section~\ref{new_systems}), as dictated by the timing of the data release.

	We have focused our companion search on late M and L dwarf companions detected in the UKIRT Infrared Deep Sky Survey (UKIDSS) Large Area Survey \citep[ULAS,][]{2007MNRAS.379.1599L} and the Sloan Digital Sky Survey \citep[SDSS,][]{2000AJ....120.1579Y}. We selected an initial photometric sample designed to contain the latest M dwarfs and L dwarfs, using basic colour cuts to exclude earlier M dwarfs and other stars \citep[e.g.][]{2010AJ....139.1808S}, while at the same time including unusual objects \citep[sub-dwarfs, metal rich/poor dwarfs, young objects; c.f.][]{2013MNRAS.430.1171D,2014ApJ...783..122K,2017MNRAS.464.3040Z}. We also imposed a signal-to-noise limit to avoid large numbers of low quality candidates. Our initial selection criteria are listed below.
 	
 	\begin{description}
 		\item{$Y - J > 0.85$~mag}
 		\item{$J - H > 0.50$~mag}
 		\item{$z - J > 2.1$~mag}
 		\item{$\sigma_J < 0.10$~mag}
 	\end{description}

 	We cross-matched our initial photometric sample with the list of possible primaries to identify potential pairings out to a maximum matching radius of 3~arcmin (following Section~\ref{priority}). For each candidate system we used the catalogued distance constraint of the primary to calculate an absolute magnitude estimate for the candidate companion (which assumes common-distance for the candidate system). We then imposed colour-absolute magnitude criteria (based on known UCDs with measured parallax), and thus excluded candidate companions whose colour-magnitude measurements were inconsistent with late M/L dwarf companionship. We constructed our known ultracool sample using the compilation of \citet{2012ApJS..201...19D} supplemented with additional objects (with parallax) from DwarfArchives\footnote{\url{http://spider.ipac.caltech.edu/staff/davy/ARCHIVE/index.shtml}}. \textit{Two Micron All-Sky Survey} \citep[2MASS,][]{2006AJ....131.1163S}, UKIDSS, and SDSS photometry was obtained where available, and selection criteria established for $M_J$~vs.~$(z-J)$ and $M_z$~vs.~$(i-z)$ colour magnitude diagrams.

\begin{figure}
 		  \includegraphics[width=0.5\textwidth]{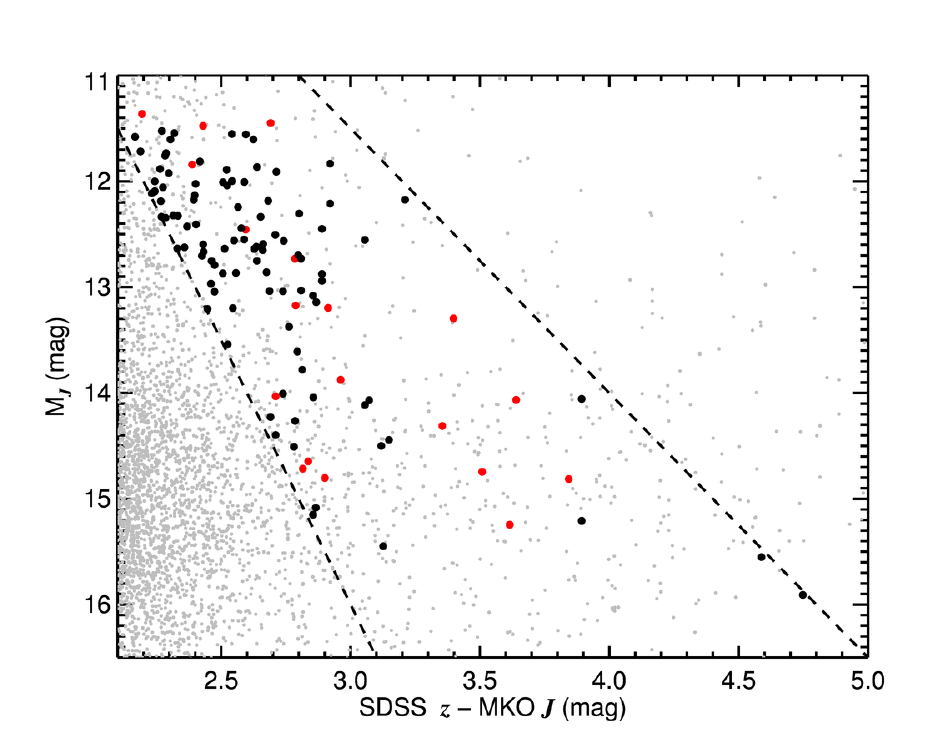}
 		  \includegraphics[width=0.5\textwidth]{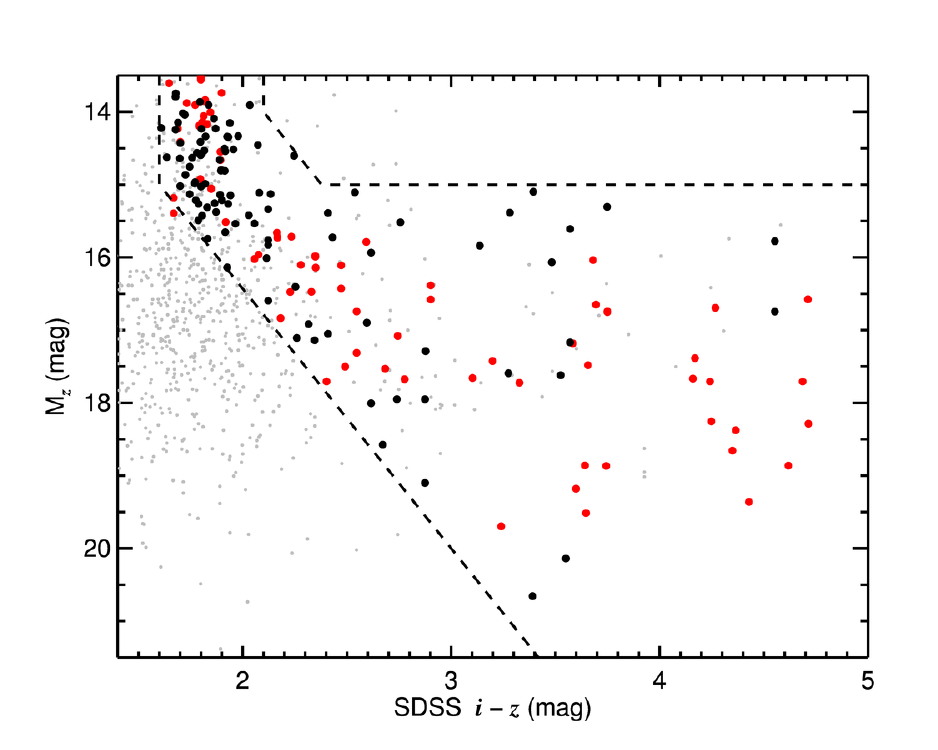}
 		  \caption{The colour-magnitude diagrams used to select our benchmark candidates. In each plot the selection region is enclosed by dashed lines, with good candidates plotted in black and rejected objects (considering both diagrams) plotted in gray. Red filled circles indicate known UCDs (with measured parallaxes) that were used to guide the selection criteria. \label{cm_plot}}
 	\end{figure}

 	The known sample includes a variety of unusual objects \citep[including sub-dwarfs, low-metallicity objects, moving group members and other young objects, several planetary mass objects, and unresolved multiples; see Table 9 of][]{2012ApJS..201...19D}. Our selection criteria should thus be reasonably inclusive, and effective at selecting companions with a range of properties while rejecting contamination. Figure~\ref{cm_plot} shows our colour-absolute magnitude criteria as dashed lines, with the known parallax sample plotted as red circles. The late M/L sequence is clear (despite some scatter), and is enclosed by the dashed lines, which are defined here:

 	\begin{description}
 		\item $[2.5 \times (z - J) + 4] < M_J < [5 \times (z - J) + 1]$ AND $M_J > 11.5$
 		\item $1.6 < (i - z) < 6.0$
 		\item $11.5 < M_z < [3.5714 \times (i - z) +9.286]$
 	\end{description}

 	\noindent We then removed objects with the following:

 	\begin{description}
 		\item $M_z < 15$ AND $M_z < [3.5714 \times (i - z) + 6.5]$ AND $(i - z) > 2.1$
 	\end{description}

	Figure~\ref{cm_plot} also shows our initial photometric sample and our selected candidates. Visual inspection of SDSS/UKIDSS images identified contamination from diffraction spikes, resolved galaxies, and some mis-matches between the SDSS and ULAS, resulting in a final sample of 100 candidate benchmark systems.

\begin{figure}
   		\includegraphics[width=0.5\textwidth]{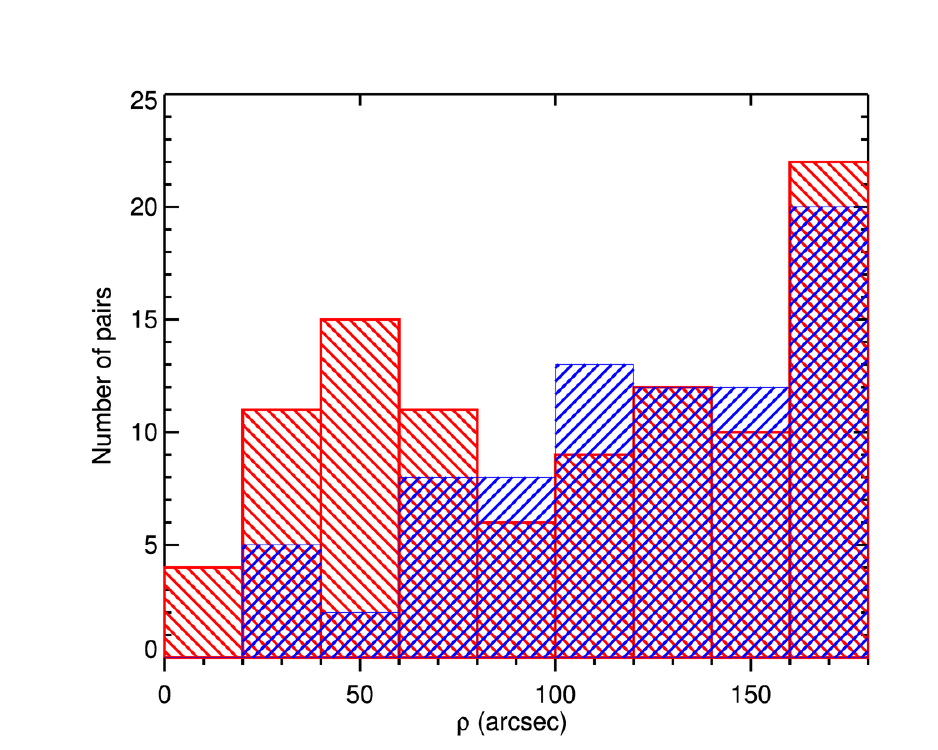}
   		\caption{The separation distribution of our candidates (in red), compared with the separation distribution for random pairs (in blue). It is clear that we are retrieving a population of real binaries with angular separation $\rho\,<\,$80~arcsec, while at larger separations spurious matches are likely to be increasingly common. \label{sep_dist}}
   	\end{figure}

	Although our selection method rules out much contamination, producing a candidate list that is rich with genuine systems, observational confirmation is still an important requirement in order to reject spurious associations. In Figure~\ref{sep_dist} we compare the separation distribution for our benchmark candidates with the separation distribution of random pairs of objects in the sky. The random pairs are generated by shifting the primary stars by 10~arcmin to the west in Galactic longitude, thus creating a ``control sample'' of stars with the same density and apparent magnitude distribution of our real sample. Any real binary within our sample would however be broken, since the shift is larger than our cross-matching radius \citep[this method is similar to the ``dancing pairs'' method described in][]{2007AJ....133..889L}. While the distribution of the ``control sample'' increases with separation (as a result of the larger area probed), the real sample shows an excess of systems out to $\rho\,\sim\,1.5$~arcmin. At larger separations, spurious matches are likely to be increasingly common. 

   	\begin{table}
   		\centering
   		\caption{The compilation of catalogues containing distance measurements and metallicity constraints from spectroscopy or narrow band photometry (or in the case of \citeauthor{2006ApJ...638.1004A} broad band photometry and proper motion) used to produce our list of possible primaries. \label{primary_summary}}
   		\begin{tabular}{l c}
   		Source & Number of stars \\
   		\hline
   		\hline
   		\citet{2006ApJ...638.1004A} & 611,798 \\
   		\citet{2010AeA...521A..40A} & 459 \\
   		\citet{2012Natur.486..375B} & 226 \\
   		\citet{2006MNRAS.373...13C} & 104 \\
   		\citet{2008MNRAS.389..585C} & 343 \\
   		\citet{1993AeA...275..101E} & 189 \\
   		\citet{2005AeA...430..165F} & 6,690 \\
   		\citet{2001AeA...377..911F} & 5,828 \\
   		\citet{2005ApJ...622.1102F} & 105 \\
   		\citet{2010ApJ...720.1290G} & 265 \\
   		\citet{2007AeA...475.1003H} & 366 \\
   		\citet{2002AeA...394..927I} & 493 \\
   		\citet{2008AeA...485..571J} & 322 \\
   		\citet{2005MNRAS.360.1345K} & 437 \\
   		\citet{2013AJ....146..134K} & 482,194 \\
   		\citet{2001ARep...45..972K} & 43 \\
   		\citet{2004MNRAS.349..757L} & 451 \\
   		\citet{2012AeA...541A..40M} & 119 \\
   		\citet{2013AeA...554A..84M} & 142 \\
   		\citet{1997AeAS..124..359M} & 146 \\
   		\citet{1995BICDS..47...13M} & 5,498 \\
   		\citet{2003ARep...47..422M} & 100 \\
   		\citet{2013AeA...551A.112M} & 1798 \\
   		\citet{2012ApJ...750L..37M} & 116 \\
   		\citet{2013AeA...551A..36N} & 254 \\
   		\citet{2003AeA...404..689N} & 54 \\
   		\citet{2004AeA...418..989N} & 16,682 \\
   		\citet{1998MNRAS.298..332R} & 730 \\
   		\citet{2012ApJ...748...93R} & 133 \\
   		\citet{2004AeA...415.1153S} & 98 \\
   		\citet{2011AeA...526A.112S} & 88 \\
   		\citet{2003AeA...398..141S} & 387 \\
   		\citet{2010AeA...515A.111S} & 64,082 \\
   		\citet{2011AeA...533A.141S} & 582 \\
   		\citet{2009AIPC.1170..255S} & 1,009 \\
   		\citet{1999AJ....118..895T} & 503 \\
   		\citet{2015MNRAS.448..855Y} & 2,207,803 \\
   		\hline
   		\end{tabular}
   	\end{table}

\section{Spectroscopic Observations} 
	\label{spectroscopy}
	We followed up 37 UCD candidates and two primaries using the Optical System for Imaging and low-Intermediate-Resolution Integrated Spectroscopy \citep[OSIRIS,][]{2003SPIE.4841.1739C} on the Gran Telescopio Canarias (GTC), and the Long-slit Intermediate Resolution Infrared Spectrograph \citep[LIRIS,][]{2003INGN....7...15A,2004SPIE.5492.1094M} on the William Herschel Telescope (WHT). Seven of the UCD candidates were found to be contaminants (reddened background stars and galaxies), while the other 30 were confirmed UCDs. Further details on the observation strategy, data reduction and analysis can be found in the following subsections, as well as in Appendix \ref{obs_log}. 

	\subsection{GTC/OSIRIS}
		OSIRIS is an imager and spectrograph for the optical wavelength range, located in the Nasmyth-B focus of GTC. We used its long-slit spectroscopy mode with the R300R grism, covering the 4800$-$10000~\AA{} range at a resolution of $\sim$350 (7.74~\AA~pix$^{-1}$). Observations were carried out in service mode, with a spectrophotometric standard observed with each group of targets. 

		The data were reduced using standard IRAF routines. Raw spectra were de-biased, and flat-fielded. We fit a low-order polynomial to remove the sky background, and then extracted the resulting spectra. Wavelength calibration was achieved with the aid of xenon-argon arc lamps, while the observed spectrophotometric standards were used for flux-calibration and first-order telluric correction.

		OSIRIS red grisms suffer from a slight contamination in the spectrum due to the second order, as the spectral order sorter filter does not block completely the contribution for wavelengths lower that the defined cut level. Hence, there is a distinguishable contamination from light coming from the 4800--4900~\AA{} range, whose second order contributes in the 9600--9800~\AA{} range, depending on the source spectral energy distribution (hereafter SED). This effect is therefore negligible for our UCD candidates (whose SED is very red) but does affect the blue stars that we used as spectrophotometric standards. To correct for that, we repeated the observations of each standard using the $z$ filter to block any second order contamination, and obtain a ``clean'' 9600--9800~\AA{} spectrum.

\begin{figure}
			\includegraphics[width=0.5\textwidth, height=0.9\textheight]{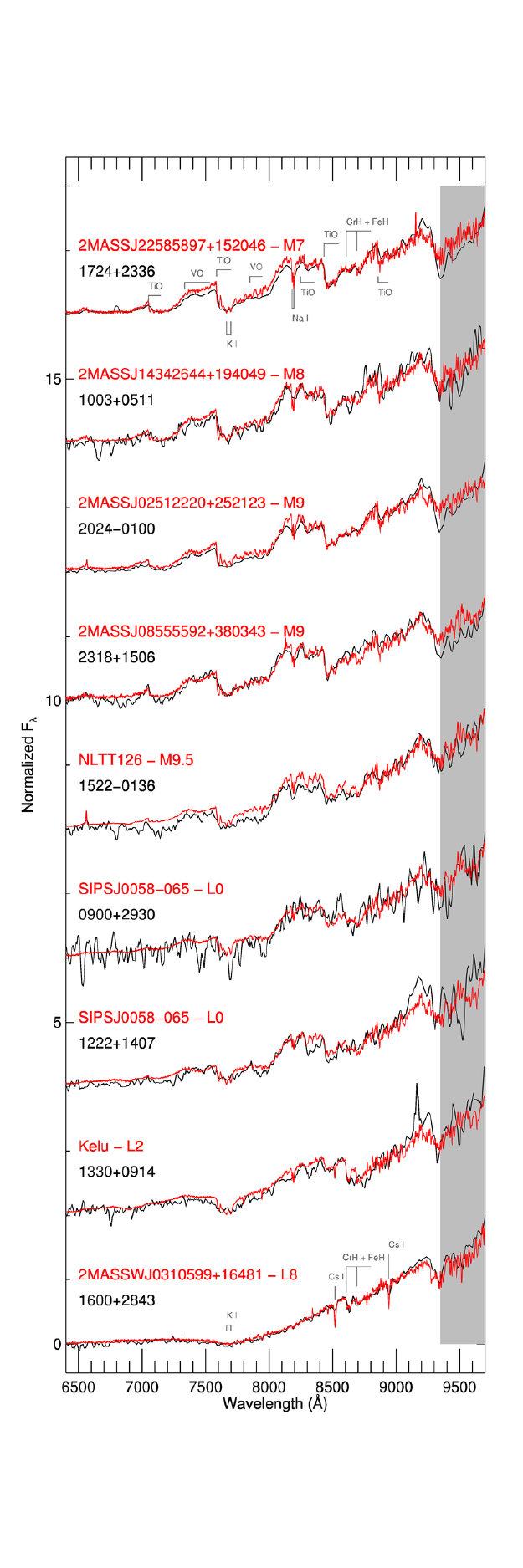}
			\caption{The GTC/OSIRIS spectra for our nine UCD candidates, sorted in ascending order of spectral type. For each UCD we overplot in red the best fit template used for spectral typing. All templates are taken from the ``Keck LRIS spectra of late-M, L and T dwarfs'' library. The grey shaded area is excluded from the fits. \label{gtc_spectra}}
		\end{figure}

		The GTC/OSIRIS spectra for the nine confirmed UCDs can be seen in Figure~\ref{gtc_spectra}.

	\subsection{WHT/LIRIS}
		LIRIS is a NIR imager and spectrograph mounted on the Cassegrain focus of the WHT. We used the long-slit spectroscopy mode with the 0.75~arcsec slit and the \emph{lr\_zj} grism, covering the 8870$-$15310~\AA{} wavelength range at a resolution of $\sim$700 (6.1~\AA{}~pix$^{-1}$). Observations were carried out in visitor mode, adopting a target-standard-target schedule to minimize overheads. We observed both target and standards following an ABBA pattern, with a dither offset of 12~arcsec. 

		IRAF routines were used to perform the standard steps of data reduction, i.e. de-biasing, flat-fielding, and pair-wise subtraction. We then median combined the individual exposures and extracted the resulting spectra. Wavelength calibration was achieved observing xenon and argon arc lamps separately, to maximize the number of available lines while avoiding saturating the strongest lines. The standards were chosen preferentially among late-B- and early-A-type stars in close proximity of our targets, to minimize the difference between the airmass of the observations. 

\begin{figure*}
			\includegraphics[width=0.49\textwidth, height=0.9\textheight]{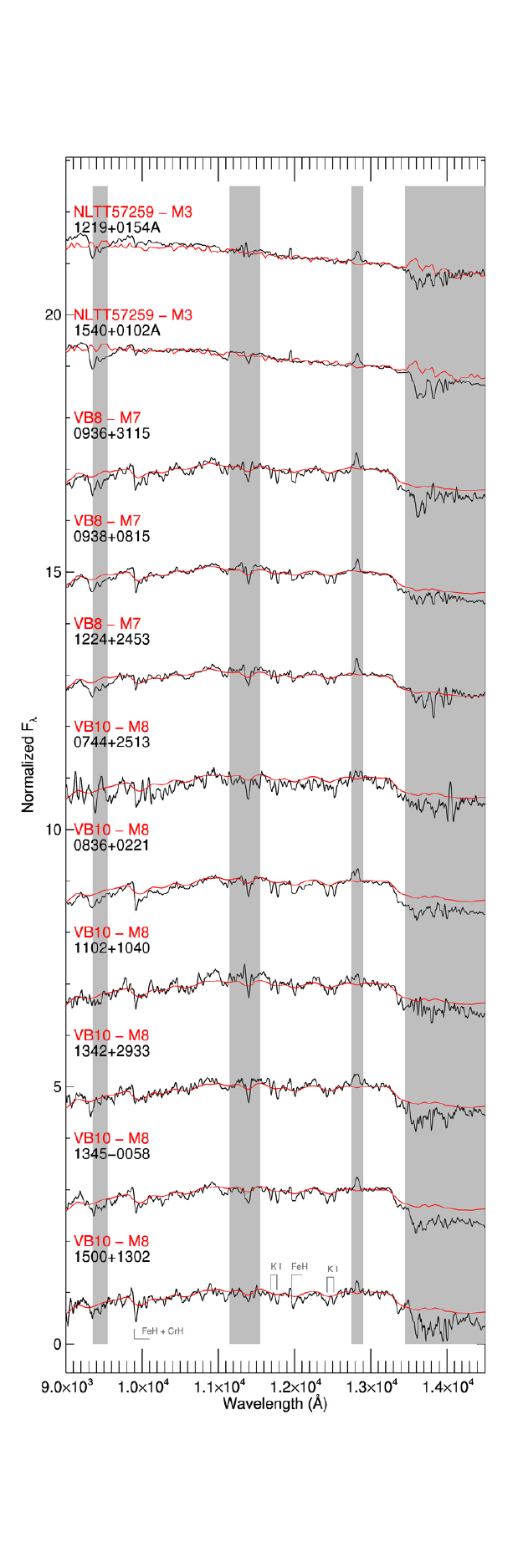}
			\includegraphics[width=0.49\textwidth, height=0.9\textheight]{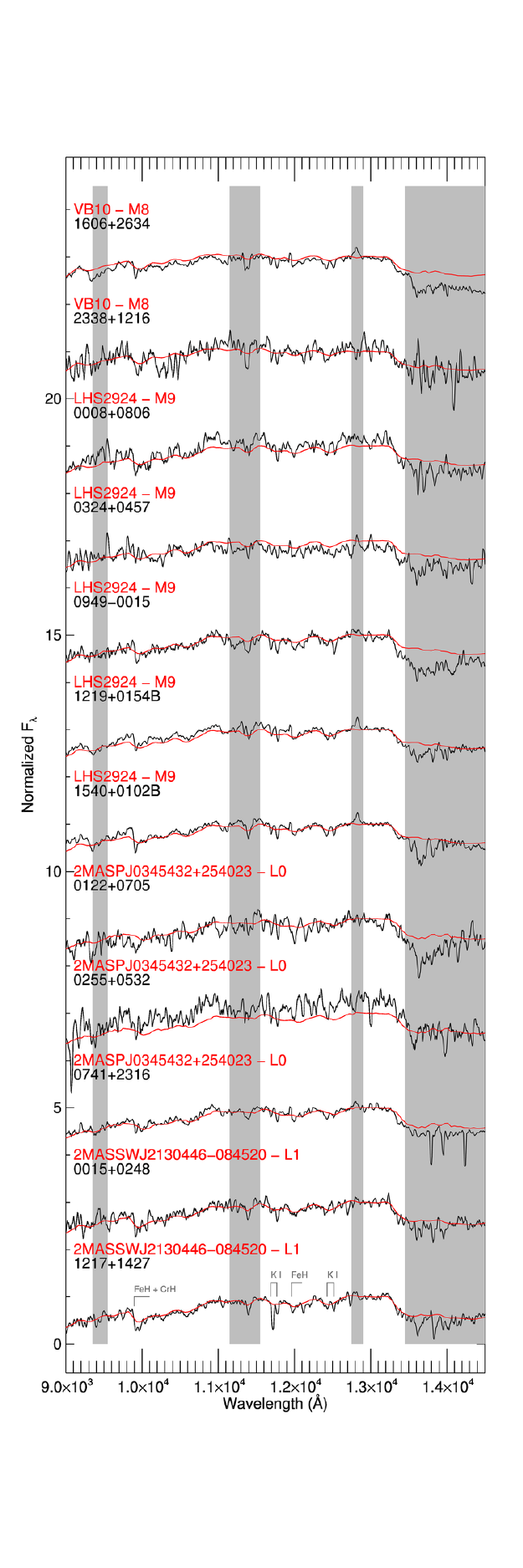}
			\caption{The WHT/LIRIS spectra for our 21 UCD candidates and two primaries, sorted by spectral type. For each UCD we overplot in red the best fit template used for spectral typing. All templates are taken from the SpeX-Prism online library. Grey shaded areas are excluded from the fits. \label{wht_spectra}}
		\end{figure*}

		The WHT/LIRIS spectra for the 21 confirmed UCDs as well as two primaries can be seen in Figure~\ref{wht_spectra}.

	\subsection{Spectral types}
		We determined spectral types for our UCDs via $\chi^2$ minimization template matching, using our own IDL routines. For the WHT/LIRIS spectra, we used the low-resolution, near-infrared templates taken from the SpeX-Prism on-line library\footnote{\url{http://www.browndwarfs.org/spexprism}}. The spectra of our targets were smoothed down to the same resolution of the SpeX-Prism templates, and we avoided the telluric bands when computing the $\chi^2$ statistic. The quality of the fits was then assessed by eye in order to identify any peculiarity in the spectra of the targets. For the GTC/OSIRIS spectra, we used the optical templates from the ``Keck LRIS spectra of late-M, L and T dwarfs''\footnote{\url{http://svo2.cab.inta-csic.es/theory/newov/templates.php?model=tpl_keck}}. The OSIRIS spectra were treated in the same way described above. We restricted the fitting to the 5,000--9,350~\AA{} range, since we noticed a systematic offset in the spectral shape between OSIRIS and LIRIS at wavelength longer than 9,350~\AA. This effect is probably to be attributed to the differences in the instrumental response functions. For each target, we overplot the best fit template in red in Figures~\ref{gtc_spectra}--\ref{wht_spectra}, while grey shaded areas are those either highly affected by telluric absorption, or by instrumental effects, and are therefore excluded from the fit.

\section{Proper motion measurements}
	\label{proper_motion}
	To contribute to the statistical assessment of the companionship of our UCDs we measured their proper motion. When available, we used the measured proper motions tabulated by \citet{2014MNRAS.437.3603S}. For objects outside the area covered by \citet{2014MNRAS.437.3603S}, we used the SDSS, ULAS, and if available the 2MASS positions to compute the proper motion. We took the catalogue coordinates and their (published) associated uncertainties, and determined the proper motion via a linear fit to $\alpha$ and $\delta$ \citep[through weighted least-square minimization; for more details, see ``Solution by Use of Singular Value Decomposition'', Section~15.4, in][]{2002nrca.book.....P}. Proper motions of the primary and secondary components are presented in Table~\ref{astrometry}, which also contains the $\chi^2$ of the linear fit. The $\chi^2$ is presented only for the values calculated here, and only for those objects detected in all three epochs (the $\chi^2$ otherwise is, by definition, zero).

	This approach leads to a very heterogeneous level of precision, depending on the time baseline covered (from more than 10~years in the best cases, to less than 1~year in the worst cases), the brightness of the object in SDSS and 2MASS (and the resulting centroiding precision), and the magnitude of the proper motion itself. Also, for the values measured here our method does not take into account possible systematic shifts between the 2MASS, SDSS, and ULAS catalogues. Our derived proper motions have precision ranging from $\sim$5~mas~yr$^{-1}$ to $\sim$50~mas~yr$^{-1}$, but for the reasons explained above these should be taken as lower limits on the real precision. 

	\onecolumn
	\begin{deluxetable}{l c c c c c c c c}

		\tablecolumns{9}
		\tablewidth{0pt}
		\tabletypesize{\scriptsize}
		
        \tablehead{
			\colhead{ID} & \colhead{$\mu_\alpha\,\cos{\delta}$} & \colhead{$\mu_\delta$} & \colhead{$\chi^2_\alpha$} & \colhead{$\chi^2_\delta$} &\colhead{$\mu$ ref.} & \colhead{$d$} & \colhead{$d$ ref.} & \colhead{Binary?} \\
			\colhead{} & \colhead{[mas~yr$^{-1}$]} & \colhead{[mas~yr$^{-1}$]} & \colhead{} & \colhead{} & \colhead{} & \colhead{[pc]} & \colhead{} & \colhead{}}

		\tablecaption{Summary of the astrometric parameters of our systems. \label{astrometry}}

        \startdata

		ULAS~J00081284+0806421 & $-41\,\pm\,20$ & $-236\,\pm\,17$ & 1.3 & 1.3 & 1 & $112\,\pm\,21$ & 1 & \multirow{2}{*}{N}\\
		BD+07~3 & $-1.3\,\pm\,1.3$ & $-19.26\,\pm\,0.60$ & \ldots & \ldots & 3 & 393$^{+80}_{-56}$ & 3 & \\
		\hline
		ULAS~J00151479+0248020 & $69\,\pm\,11$ & $8\,\pm\,11$ & 3.6 & 0.010 & 1 & $62\,\pm\,12$ & 1 & \multirow{2}{*}{R}\\
		2MASS~J00151561+0247373 & $81.8\,\pm\,1.9$ & $-2.6\,\pm\,1.3$ & \ldots & \ldots & 4 & $61\,\pm\,20$ & 1 & \\
		\hline
		ULAS~J01223706+0705579 & $-68\,\pm\,42$ & $-18\,\pm\,40$ & \ldots & \ldots & 1 & $130\,\pm\,24$ & 1 & \multirow{2}{*}{N}\\
		TYC~27-721-1 & $8.8\,\pm\,2.0$ & $-25.80\,\pm\,0.48$ & \ldots & \ldots & 3 & $584^{+135}_{-92}$ & 3 & \\
		\hline
		ULAS~J02553253+0532122 & $28\,\pm\,30$ & $40\,\pm\,30$ & \ldots & \ldots & 1 & $140\,\pm\,26$ & 1 & \multirow{2}{*}{U}\\
		TYC~54-833-1 & $-13.83\,\pm\,0.14$ & $-2.01\,\pm\,0.15$ & \ldots & \ldots & 3 & $174^{+9}_{-8}$ & 3 & \\
		\hline
		ULAS~J03244133+0457520 & $21\,\pm\,11$ & $-34\,\pm\,11$ & 1.6 & 3.2 & 1 & $124\,\pm\,23$ & 1 & \multirow{2}{*}{N}\\
		BD+04~533 & $49.64\,\pm\,0.64$ & $-13.60\,\pm\,0.39$ & \ldots & \ldots & 3 & $546^{+83}_{-63}$ & 3 & \\
		\hline
		ULAS~J07410439+2316376 & $-18.8\,\pm\,5.5$ & $12.3\,\pm\,5.3$ & 2.2 & 18 & 1 & $80\,\pm\,15$ & 1 & \multirow{2}{*}{N}\\
		TYC~1912-724-1 & $-9.4\,\pm\,1.7$ & $-12.6\,\pm\,1.6$ & \ldots & \ldots & 3 & $527^{+83}_{-63}$ & 3 & \\
		\hline
		ULAS~J07443600+2513306 & $-41.4\,\pm\,7.4$ & $-10.3\,\pm\,5.7$ & \ldots & \ldots & 2 & $132\,\pm\,26$ & 1 & \multirow{2}{*}{N}\\
		TYC~1916-1611-1 & $4.2\,\pm\,1.2$ & $-7.2\,\pm\,1.2$ & \ldots & \ldots & 3 & $359^{+39}_{-32}$ & 3 & \\
		\hline
		ULAS~J08361347+0221063 & $-100.3\,\pm\,4.6$ & $33.5\,\pm\,5.2$ & \ldots & \ldots & 2 & $53\,\pm\,10$ & 1 & \multirow{2}{*}{N}\\
		BD+02~2020 & $7.9\,\pm\,2.1$ & $-25.6\,\pm\,2.9$ & \ldots & \ldots & 3 & $476^{+262}_{-125}$ & 3 & \\
		\hline
		ULAS~J09000474+2930221 & $-13\,\pm\,10$ & $-27.8\,\pm\,8.8$ & \ldots & \ldots & 2 & $197\,\pm\,37$ & 1 & \multirow{2}{*}{R}\\ 
		NJCM~J09001350+2931203 & $-9.7\,\pm\,3.8$ & $-17.0\,\pm\,3.8$ & \ldots & \ldots & 5 & $123\,\pm\,20$ & 5 & \\
		\hline
		ULAS~J09361316+3115135 & $-151.5\,\pm\,4.7$ & $26.7\,\pm\,4.7$ & \ldots & \ldots & 2 & $155\,\pm\,35$ & 1 & \multirow{2}{*}{N}\\
		NJCM~J09361658+3116368 & $31.6\,\pm\,3.8$ & $-88.9\,\pm\,3.8$ & \ldots & \ldots & 5 & $58\,\pm\,20$ & 5 & \\
		\hline
		ULAS~J09383678+0815110 & $-116.3\,\pm\,6.0$ & $-44.4\,\pm\,5.2$ & 0.0013 & 1.3 & 1 & $55\,\pm\,13$ & 1 & \multirow{2}{*}{N}\\
		TYC~821-1173-1 & $5.4\,\pm\,2.9$ & $-22.8\,\pm\,1.4$ & \ldots & \ldots & 3 & $199^{+36}_{-26}$ & 3 & \\
		\hline
		ULAS~J09493641$-$0015334 & $155\,\pm\,17$ & $-84\,\pm\,17$ & 1.008 & 0.18 & 1 & $76\,\pm\,14$ & 1 & \multirow{2}{*}{N}\\
		IDS~09445+0011~AB & $-39.6\,\pm\,0.9$ & $19.3\,\pm\,1.1$ & \ldots & \ldots & 4 & $55\,\pm\,21$ & 1 & \\
		\hline
		ULAS~J10033792+0511417 & $-9\,\pm\,10$ & $-28.2\,\pm\,9.9$ & \ldots & \ldots & 2 & $272\,\pm\,53$ & 1 & \multirow{2}{*}{N}\\
		BD+05~2275 & $-11.1\,\pm\,1.7$ & $5.97\,\pm\,0.92$ & \ldots & \ldots & 3 & $223^{+26}_{-21}$ & 3 & \\
		\hline
		ULAS~J11025103+1040466 & $-132\,\pm\,12$ & $-41\,\pm\,11$ & 0.12 & 0.0067 & 1 & $122\,\pm\,24$ & 1 & \multirow{2}{*}{N}\\
		2MASS~J11025520+1041036 & $-71.8\,\pm\,3.3$ & $-81.5\,\pm\,3.1$ & \ldots & \ldots & 4 & $96\,\pm\,20$ & 1 & \\
		\hline
		ULAS~J12173673+1427096 & $-74\,\pm\,20$ & $-34\,\pm\,20$ & 3.0 & 1.5 & 1 & $70\,\pm\,13$ & 1 & \multirow{2}{*}{R}\\
		HD~106888 & $-102.589\,\pm\,0.047$ & $-37.669\,\pm\,0.031$ & \ldots & \ldots & 3 & $74^{+4}_{-3}$ & 3 & \\
		\hline
		ULAS~J12193254+0154330 & $-72.9\,\pm\,5.9$ & $-82.1\,\pm\,5.9$ & 2.8 & 0.80 & 1 & $54\,\pm\,10$ & 1 & \multirow{2}{*}{R}\\
		PYC~12195+0154 & $-73.4\,\pm\,4.7$ & $-66.1\,\pm\,5.4$ & \ldots & \ldots & 4 & $67\,\pm\,15$ & 1 & \\
		\hline
		ULAS~J12225930+1407501 & $-49.0\,\pm\,8.2$ & $-19.7\,\pm\,8.5$ & \ldots & \ldots & 2 & $216\,\pm\,41$ & 1 & \multirow{2}{*}{R}\\
		NJCM~J12225728+1407185 & $-43.9\,\pm\,4.1$ & $-10.9\,\pm\,4.1$ & \ldots & \ldots & 5 & $155\,\pm\,20$ & 5 & \\
		\hline
		ULAS~J12241699+2453334 & $-30.4\,\pm\,1.7$ & $-49.4\,\pm\,1.8$ & 2.2 & 1.8 & 1 & $86\,\pm\,19$ & 1 & \multirow{2}{*}{N}\\
		TYC~1989-265-1 & $-11.5\,\pm\,1.4$ & $-4.8\,\pm\,1.0$ & \ldots & \ldots & 3 & $319^{+79}_{-52}$ & 3 & \\ 
		\hline
		ULAS~J13300249+0914321 & $-83\,\pm\,37$ & $10\,\pm\,37$ & \ldots & \ldots & 1 & $149\,\pm\,30$ & 1 & \multirow{2}{*}{R}\\
		TYC~892-36-1 & $-9.6\,\pm\,2.7$ & $11.37\,\pm\,0.40$ & \ldots & \ldots & 3 & $233^{+27}_{-22}$ & 3 & \\ 
		\hline
		ULAS~J13420199+2933400 & $-99\,\pm\,14$ & $53\,\pm\,12$ & 0.027 & 0.0022 & 1 & $96\,\pm\,19$ & 1 & \multirow{2}{*}{N}\\
		BD+30~2436 & $-47.89\,\pm\,0.53$ & $-8.92\,\pm\,0.38$ & \ldots & \ldots & 3 & $60\,\pm\,20^a$ & 1 & \\
		\hline
		ULAS~J13451242$-$0058443 & $-66\,\pm\,16$ & $15\,\pm\,16$ & 0.15 & 0.33 & 1 & $107\,\pm\,21$ & 1 & \multirow{2}{*}{N}\\
		NJCM~J13451873$-$0057295 & $-22.5\,\pm\,5.0$ & $-53.3\,\pm\,5.0$ & \ldots & \ldots & 5 & $75\,\pm\,16$ & 5 & \\
		\hline
		ULAS~J15001074+1302122 & $30\,\pm\,14$ & $-31\,\pm\,14$ & 0.82 & 0.51 & 1 & $138\,\pm\,27$ & 1 & \multirow{2}{*}{N}\\
		HD~132681 & $0.6\,\pm\,1.4$ & $11.94\,\pm\,0.85$ & \ldots & \ldots & 3 & $269^{+22}_{-19}$ & 3 & \\ 
		\hline
		ULAS~J15224658$-$0136426 & $-53\,\pm\,20$ & $-7\,\pm\,20$ & \ldots & \ldots & 1 & $227\,\pm\,43$ & 1 & \multirow{2}{*}{R}\\
		HIP~75262 & $-67.55\,\pm\,0.65$ & $11.49\,\pm\,0.39$ & \ldots & \ldots & 3 & $210^{+11}_{-9}$ & 3 & \\ 
		\hline
		ULAS~J15400510+0102088 & $-50.3\,\pm\,7.4$ & $-0.8\,\pm\,5.8$ & 0.0015 & 0.22 & 1 & $58\,\pm\,11$ & 1 & \multirow{2}{*}{R}\\
		NJCM~J15400591+0102151 & $-39.6\,\pm\,3.6$ & $-8.4\,\pm\,3.6$ & \ldots & \ldots & 5 & $44\,\pm\,21$ & 5 & \\
		\hline
		ULAS~J16003655+2843062 & $-222.7\,\pm\,5.6$ & $230.7\,\pm\,5.6$ & \ldots & \ldots & 2 & $44.3\,\pm\,8.2$ & 1 & \multirow{2}{*}{N}\\
		TYC~2041-1324-1 & $-8.64\,\pm\,0.78$ & $-5.2\,\pm\,1.3$ & \ldots & \ldots & 3 & $370^{+32}_{-27}$ & 3 & \\ 
		\hline
		ULAS~J16061153+2634518 & $-50.7\,\pm\,6.3$ & $132.5\,\pm\,8.8$ & \ldots & \ldots & 2 & $91\,\pm\,18$ & 1 & \multirow{2}{*}{R}\\
		TYC~2038-524-1 & $-43.99\,\pm\,0.42$ & $141.16\,\pm\,0.72$ & \ldots & \ldots & 3 & $116^{+3}_{-3}$ & 3 & \\  
		\hline
		SDSS~J172437.52+233649.3 & $-224\,\pm\,47$ & $94\,\pm\,47$ & \ldots & \ldots & 1 & $101\,\pm\,23$ & 1 & \multirow{2}{*}{N}\\
		TYC~2074-442-1 & $-3.85\,\pm\,0.71$ & $5.34\,\pm\,0.67$ & \ldots & \ldots & 3 & $520^{+140}_{-91}$ & 3 & \\ 
		\hline
		SDSS~J202410.30$-$010039.2 & $-19\,\pm\,46$ & $-79\,\pm\,46$ & \ldots & \ldots & 1 & $136\,\pm\,28$ & 1 & \multirow{2}{*}{U}\\
		BD-01~3972 & $-11.49\,\pm\,0.78$ & $-18.70\,\pm\,0.67$ & \ldots & \ldots & 3 & $198^{+13}_{-11}$ & 3 & \\  
		\hline
		ULAS~J23180626+1506100 & $47\,\pm\,24$ & $38\,\pm\,24$ & \ldots & \ldots & 1 & $216\,\pm\,41$ & 1 & \multirow{2}{*}{U}\\
		2MASS~J23181098+1503259 & $9.4\,\pm\,1.7$ & $-7.1\,\pm\,2.2$ & \ldots & \ldots & 4 & $217\,\pm\,15$ & 1 & \\ 
		\hline
		ULAS~J23383981+1216341 & $35\,\pm\,33$ & $93\,\pm\,32$ & 0.031 & 2.7 & 1 & $148\,\pm\,29$ & 1 & \multirow{2}{*}{R}\\
		TYC~1172-357-1 & $28.5\,\pm\,1.9$ & $11.97\,\pm\,0.76$ & \ldots & \ldots & 3 & $477^{+275}_{-128}$ & 3 & 
		\enddata

        \tablecomments{For each system we list the UCD first and the primary second. For each object we list the two components of its proper motion, along with the source of the measurement. For the values determined in this paper, and with at least three usable epochs, we present the $\chi^2$ of the linear fit. $^a$-- TGAS reports a parallax of $0.02\,\pm\,0.29$~mas, so we chose to adopt its spectrophotometric distance instead. ``R'' stands for ``robust system'', ``U'' for ``uncertain system'', and ``N'' for ``not a binary''. References: 1 -- this paper. 2 -- \citet{2014MNRAS.437.3603S}. 3 -- \citet{2016A&A...595A...4L}. 4 -- \citet{2013AJ....145...44Z}. 5 -- \citet{2016MNRAS.457.2192C}.}
	\end{deluxetable}
	\twocolumn

\section{New benchmark systems}
	\label{new_systems}
	We have identified 13 new common distance, common proper motion systems containing a UCD companion. 

\begin{figure*}
		\includegraphics[width=0.49\textwidth]{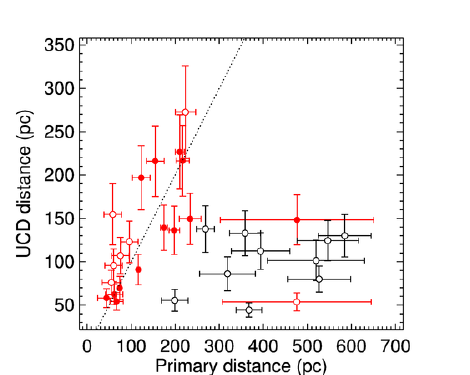}
		\includegraphics[width=0.49\textwidth]{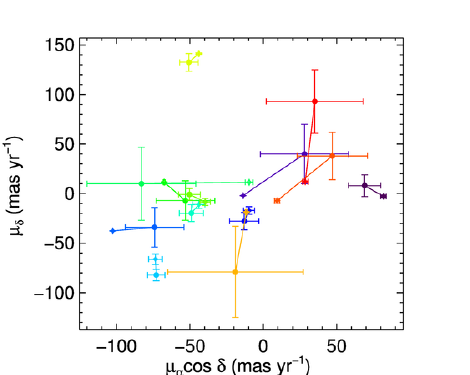}
		\caption{\emph{Left}: distance to our potential primaries against distance to their potential companions. Pairs we select as having common distance are highlighted in red. Systems that passed a subsequent common-proper-motion test are plotted as filled circles, while those that did not are plotted as open circles. The dashed line represents the one-to-one correspondence. \emph{Right}: vector point diagram for our candidate common proper motion systems. Objects that we select as common-proper motion are plotted in colour, with a line joining the primary (plotted with a star symbol) and its UCD companion (plotted with a filled circle). \label{d_pm}}
	\end{figure*}

	To ensure common distance we used, when available, published astrometric measurements. For the candidate primaries, 21 out of 30 are found in TGAS \citep{2016A&A...595A...2G,2016A&A...595A...4L}. For the remaining nine candidate primaries, we relied on spectrophotometric distance estimates, calculated following \citet{1992adps.book.....L} and using published $B$, $V$, and $I$ photometry. For the UCDs, we used the MKO $J$ spectrophotometric distance calibration presented in \citet{2012ApJS..201...19D}, taking into account photometric and spectral typing uncertainties, as well as the scatter around the published polynomial relation. We define as ``common distance pair'' only those systems where the distances agree at the $3\,\sigma$ level. Although we used $2\,\sigma$ uncertainties in our CGB simulation analysis, in practice we adopt a more liberal approach for passing candidates at each follow-up stage. In our final analysis we will still impose our desired false-alarm probability requirements, but will reduce the likelihood of ruling out systems whose spectral type and distance uncertainties may be somewhat under-estimated. Our adopted distances are summarized in Table~\ref{astrometry}, as well as in the left panel of Figure~\ref{d_pm}, where we show the distance to our potential primaries against the distance to their potential companions. In the majority of cases, the uncertainties on the UCD distance dominate. 

	We then used the measured proper motions (see Section~\ref{proper_motion}) to identify likely common proper motion (hereafter CPM) pairs. To account for the heterogeneous level of precision on the available proper motions, we adopted liberal CPM criteria. We rule out as non-CPM only those pairs whose proper motions are discrepant by $\geq\,3\,\sigma_\mu$, where $\sigma_\mu$ is the combined uncertainty on the proper motion of both components of the system (usually dominated by the UCD). We show in the right panel of Figure~\ref{d_pm} the vector point diagram for our candidate CPM systems.

	We calculated a false alarm probability for each system in two different ways. One follows the ``Confirmable \emph{Gaia} benchmarks'' analysis described in Section~\ref{prob}, where we search for mimics of our identified CPM pairs within our simulated field population. For each of our newly discovered CPM pair we simulated the field UCD population around the pair, and searched for mimics i.e. simulated UCDs that have distance and proper motion within 3~$\sigma$ of the distance and proper motion of the primary. Once again we used 3~$\sigma$ to be consistent with our follow-up rejection criteria, and note that our simulated false alarm probabilities will thus be somewhat greater (than in the 2~$\sigma$ case). However, as before the CGB status of a CPM pair was assessed by counting in how many of the 10,000 runs we found at least one mimic. We did not assess radial velocity consistency since we do not have measured radial velocities for any of our UCDs. We will refer to this approach as ``Method 1''.

	The other method does not rely on our simulations, but on the observed field population of stars from TGAS. For each candidate pair we searched for all field stars in a radius of 2~deg from the UCD, in the TGAS catalogue (since the majority of our primaries are in TGAS). The 2~degrees radius represents a compromise between the need to have a statistically significant sample of field stars, and the need for the sample to be homogeneous. We used this sample to determine the distance and proper motion distribution of the field population. The distance and proper motion distribution were essentially treated as a probability density function, which we reconstructed using kernel density estimation. We then draw 10,000 samples of stars from the reconstructed probability density function, and determined how many mimics of our system were generated. The false alarm probability was assumed to be the number of mimics divided by 10,000. If this probability is below 0.0027 (3~$\sigma$) we consider the pair to be a ``robust'' common proper motion system. Systems with larger false alarm probability are ruled out. We will refer to this approach as ``Method 2''.

	The two methods agree most of the time, i.e. if an object is a non-CGB it will also have a large false alarm probability. There are however three notable exceptions. These are systems that we consider to be real companions, and that have low false alarm probabilities according to Method 2, but that are classified as non-CGBs by Method 1. These are the systems including ULAS~J02553253+0532122, SDSS~J202410.30-010039.2, and ULAS~J23180626+1506100. As can be seen from Table~\ref{astrometry}, they are characterized by relatively low proper motions ($\sim\,50-80$~mas~yr$^{-1}$) with very large uncertainties ($\sim\,25-45$~mas~yr$^{-1}$). As a result the false alarm probability is high (with both methods), but in one case slightly below the threshold. We refer to these as ``uncertain systems'' (and label them ``U'' in column ``Binary?'' of Table~\ref{newsystems}; see Section~\ref{uncertain_pairs}) to distinguish them from those that have low false alarm probability according to both methods (labelled ``R''), and those that do not pass either test (labelled ``N'').

	Further details on each new system can be found in the following subsections, and are summarised in Table~\ref{newsystems}. 

	\subsection{``Robust'' common proper motion systems}

		\subsubsection{2MASS~J00151561+0247373 + ULAS~J00151483+0248039}
			The primary is a slightly metal poor K7 dwarf from the LAMOST DR2 ([Fe/H]$\,=\,-0.155\,\pm\,0.065$). It was originally classified as an M dwarf candidate by \citet{2013MNRAS.435.2161F}. It is too faint to be in TGAS so we had to estimate its spectrophotometric distance, which is $61\,\pm\,20$~pc. We got its proper motion from The Fourth US Naval Observatory CCD Astrograph Catalog \citep[UCAC4,][]{2013AJ....145...44Z}, $\mu_\alpha \cos{\delta}\,=\,81.8\,\pm\,1.9$~mas~yr$^{-1}$ and $\mu_\delta\,=\,-2.6\,\pm\,1.3$~mas~yr$^{-1}$. The companion is an L1 dwarf, at a spectrophotometric distance of $62\,\pm\,12$~pc and with a proper motion of $\mu_\alpha \cos{\delta}\,=\,69\,\pm\,11$~mas~yr$^{-1}$ and $\mu_\delta\,=\,8\,\pm\,11$~mas~yr$^{-1}$, measured fitting its 2MASS, SDSS and ULAS coordinates. The false alarm probability for this pair is $1\,\times\,10^{-5}$.

		\subsubsection{NJCM~J09001350+2931203 + ULAS~J09000474+2930221}
			This system is composed of an M3.5 from the NJCM catalogue and an L0 for which we obtained a GTC/OSIRIS spectrum, presented in Figure~\ref{gtc_spectra}. The primary is at a spectrophotometric distance of $123\,\pm\,20$~pc with a proper motion $\mu_\alpha \cos{\delta}\,=\,-9.7\,\pm\,3.8$~mas~yr$^{-1}$ and $\mu_\delta\,=\,-17.0\,\pm\,3.8$~mas~yr$^{-1}$. Using the method described in \citet{2013AeA...551A..36N} we obtain a metallicity of [Fe/H]$\,=\,-0.23\,\pm\,0.17$~dex. The companion is at a spectrophotometric distance of $197\,\pm\,37$~pc. Its measured proper motion is $\mu_\alpha \cos{\delta}\,=\,-13\,\pm\,10$~mas~yr$^{-1}$ and $\mu_\delta\,=\,-27.8\,\pm\,8.8$~mas~yr$^{-1}$. The false alarm probability for this pair is $4\,\times\,10^{-5}$. 

	    \subsubsection{HD~106888 + ULAS~J12173643+1427117}
	    	The primary is an F8 at a distance of $74^{+4}_{-3}$~pc, with a proper motion of $\mu_\alpha \cos{\delta}\,=\,-102.589\,\pm\,0.047$~mas~yr$^{-1}$ and $\mu_\delta\,=\,-37.669\,\pm\,0.031$~mas~yr$^{-1}$ \citep{2016A&A...595A...4L}. The primary has a metallicity of [Fe/H]$\,=\,-0.06$~dex \citep[no uncertainty given]{1995BICDS..47...13M}. The companion was observed with WHT/LIRIS and classified as a L1 based on template comparison. The proper motion for the companion was measured from a fit to its 2MASS, SDSS, and UKIDSS positions and we obtained $\mu_\alpha \cos{\delta}\,=\,-74\,\pm\,20$~mas~yr$^{-1}$ and $\mu_\delta\,=\,-34\,\pm\,20$~mas~yr$^{-1}$. The spectrophotometric distance to the companion is $69\,\pm\,13$~pc. With an angular separation of $\sim\,38''$, the false alarm probability for the pair is $9\,\times\,10^{-8}$. This system has been previously reported by \citet{2014ApJ...792..119D}.

		\subsubsection{PYC~J12195+0154 + ULAS~J12193254+0154330}
			The primary was proposed as a low-mass member of the AB~Dor moving group by \citet{2012AJ....143...80S}, however with only a low likelihood. UCAC4 \citep{2013AJ....145...44Z} reports a proper motion $\mu_\alpha \cos{\delta}\,=\,-73.4\,\pm\,4.7$~mas~yr$^{-1}$ and $\mu_\delta\,=\,-66.1\,\pm\,5.4$~mas~yr$^{-1}$. Using the Bayesian Analysis for Nearby Young AssociatioNs II \citep[BANYAN II,][]{2014ApJ...783..121G,2013ApJ...762...88M} online tool\footnote{\url{http://www.astro.umontreal.ca/~gagne/banyanII.php}} we obtain a 0$\%$ probability for the object to be a member of the AB~Dor moving group, a 14.7$\%$ probability for it to be part of the ``young field'' population (i.e. to be younger than 1~Gyr), and a 85.3$\%$ probability to be older than 1~Gyr. ULAS~J12193254+0154330 proper motion was measured fitting the 2MASS, SDSS, and ULAS coordinates, obtaining $\mu_\alpha \cos{\delta}\,=\,-72.9\,\pm\,5.9$~mas~yr$^{-1}$ and $\mu_\delta\,=\,-82.1\,\pm\,5.9$~mas~yr$^{-1}$, in $2\,\sigma$ agreement with PYC~J12195+0154 proper motion. 

			We did not find published spectra for PYC~J12195+0154 and ULAS~J12193254+0154330 so we observed both with LIRIS. Their spectra can be found in Figure~\ref{wht_spectra}. We classify PYC~J12195+0154 as M3.0V and ULAS~J12193254+0154330 as M9V based on a comparison to spectra templates taken from the previously mentioned SpeX-Prism library. The spectrophotometric distance of the two sources agree at the $2.3\sigma$ level. 

			The angular separation between the two objects is 11'', which at the average distance of the pair correspond to a projected separation $a\,\sim\,690$~au. Therefore the false alarm probability is $4\,\times\,10^{-10}$.

		\subsubsection{NJCM~J12225728+1407185 + ULAS~J12225930+1407501}
			This system is composed of an M4 from the NJCM catalogue and an L0 whose GTC/OSIRIS spectrum is presented in Figure~\ref{gtc_spectra}. The primary is at a spectrophotometric distance of $155\,\pm\,20$~pc with a proper motion $\mu_\alpha \cos{\delta}\,=\,-43.9\,\pm\,4.1$~mas~yr$^{-1}$ and $\mu_\delta\,=\,-10.9\,\pm\,4.1$~mas~yr$^{-1}$. Using the method described in \citet{2013AeA...551A..36N} we obtain a metallicity of [Fe/H]$\,=\,-0.19\,\pm\,0.17$~dex. The companion is at a spectrophotometric distance of $216\,\pm\,41$~pc. Its measured proper motion is $\mu_\alpha \cos{\delta}\,=\,-49\,\pm\,8.2$~mas~yr$^{-1}$ and $\mu_\delta\,=\,-19.7\,\pm\,8.5$~mas~yr$^{-1}$. The false alarm probability for this pair is $4\,\times\,10^{-7}$. 

		\subsubsection{TYC~892-36-1 + ULAS~J13300249+0914321}
			The primary is a K-dwarf ($B-V\,=\,0.96$~mag) from the Tycho catalogue, at a distance of $233^{+28}_{-22}$ \citep{2016A&A...595A...2G,2016A&A...595A...4L}. The photometric metallicity from \citet{2006ApJ...638.1004A} is [Fe/H]$\,=\,1.63\,\pm\,2.47$~dex. Its proper motion, taken from TGAS, is $\mu_\alpha \cos{\delta}\,=\,-9.6\,\pm\,2.6$~mas~yr$^{-1}$ and $\mu_\delta\,=\,11.37\,\pm\,0.40$~mas~yr$^{-1}$. The astrometric parameters for its companion are consistent only at the $\sim\,2.5\,\sigma$ level, mostly owing to the large uncertainties. The proper motion of the L2 companion is $\mu_\alpha \cos{\delta}\,=\,-83\,\pm\,37$~mas~yr$^{-1}$ and $\mu_\delta\,=\,10\,\pm\,37$~mas~yr$^{-1}$. Its spectrophotometric distance is $149\,\pm\,30$~pc. The false alarm probability for this pair is $2.8\,\times\,10^{-4}$, among the highest between our newly found pairs.

		\subsubsection{HIP~75262 + ULAS~J15224658-0136426}
			The primary is a slightly metal poor F5 dwarf, with measured [Fe/H]$\,=\,-0.39\,\pm\,0.11$~dex (from RAVE DR4), at a distance of $210^{+11}_{-10}$~pc \citep{2016A&A...595A...4L}. Its proper motion is $\mu_\alpha \cos{\delta}\,=\,-67.55\,\pm\,0.65$~mas~yr$^{-1}$ and $\mu_\delta\,=\,11.49\,\pm\,0.39$~mas~yr$^{-1}$. The companion is an M9.5 at a spectrophotometric distance of $227\,\pm\,43$~pc, and its measured proper motion is $\mu_\alpha \cos{\delta}\,=\,-53\,\pm\,20$~mas~yr$^{-1}$ and $\mu_\delta\,=\,-7\,\pm\,20$~mas~yr$^{-1}$. Distance and proper motion are therefore in good agreement, but the false alarm of the pair is $3\,\times\,10^{-4}$, the highest of our sample. This is mostly due to the significant separation between the two components, $270''$, corresponding to $\sim\,60,000$~au. 

			We note that the primary is reported by The Washington Double Star Catalog \citep{1997A&AS..125..523W} to be a companion to HIP~75261 (a K3III). However the parallaxes and proper motion of the two stars from TGAS are inconsistent with each other.

		\subsubsection{NJCM~J15400591+0102150 + ULAS~J15400510+0102088}
			NJCM~J15400591+0102150 is a M dwarf identified in NJCM. The low-resolution NIR spectrum obtained with LIRIS (presented in Figure~\ref{wht_spectra}) shows this object is a M3.0V. Its PPMXL proper motion is $\mu_\alpha \cos{\delta}\,=\,-39.6\,\pm\,3.6$~mas~yr$^{-1}$ and $\mu_\delta\,=\,-8.4\,\pm\,3.6$~mas~yr$^{-1}$. We have estimated its photometric metallicity using the method described in \citet{2013AeA...551A..36N}, and obtained [Fe/H]$\,=\,-0.14\,\pm\,0.17$~dex. The proper motion for ULAS~J15400510+0102088 obtained fitting its 2MASS, SDSS, and ULAS positions, is $\mu_\alpha \cos{\delta}\,=\,-50.3\,\pm\,7.4$~mas~yr$^{-1}$ and $\mu_\delta\,=\,-0.8\,\pm\,5.8$~mas~yr$^{-1}$. The LIRIS spectrum of ULAS~J15400510+0102088 can be found in Figure~\ref{wht_spectra} and we classify this object M9V. The spectrophotometric distance of NJCM~J15400591+0102150 and ULAS~J15400510+0102088 is $87\,\pm\,5$~pc and $60\,\pm\,12$~pc respectively, in good agreement with each other ($2.1\sigma$). The angular separation of the pair is only 13.6'', corresponding to a projected separation $a\,\sim\,910$~au. The false alarm probability is $2\,\times\,10^{-9}$.

		\subsubsection{TYC~2038-524-1 + ULAS~J16061153+2634518}
			The primary is a metal poor late-G type star ([Fe/H]$\,=\,-0.48\,\pm\,0.8$~dex; \citealt{2006ApJ...638.1004A}) from the Tycho catalogue. Its proper motion is $\mu_\alpha \cos{\delta}\,=\,-42.6\,\pm\,1.2$~mas~yr$^{-1}$ and $\mu_\delta\,=\,140.3\,\pm\,1.8$~mas~yr$^{-1}$, in very good agreement with the proper motion of the UCD, which we obtained fitting the 2MASS, SDSS, and ULAS coordinates of the target. The proper motion of the UCD is $\mu_\alpha \cos{\delta}\,=\,-50.7\,\pm\,6.3$~mas~yr$^{-1}$ and $\mu_\delta\,=\,132.5\,\pm\,8.8$~mas~yr$^{-1}$. The spectrum of ULAS~J16061153+2634518 can be found in Figure~\ref{wht_spectra}. We classify it as a M8V based on the comparison with the spectroscopic standards taken from the SpeX-Prism online library\footnote{\url{http://www.browndwarfs.org/spexprism}}. The distance of the two sources agree at the $1\sigma$ level, with TYC~2038-524-1 being at $116\,\pm\,3$~pc \citep{2016A&A...595A...2G,2016A&A...595A...4L} and ULAS~J16061153+2634518 being at $110\,\pm\,25$~pc \citep[using the polynomial relation from][]{2012ApJS..201...19D}. The two objects are separated by 50.74'', which at the average distance of the pair correspond to a projected separation $a\,\sim\,4550$~au. Therefore the false alarm probability is $8\,\times\,10^{-11}$.

		\subsubsection{TYC~1172-357-1 + ULAS~J23383981+1216341}
			The primary is an F type star, at a distance of $476^{+276}_{-127}$ (TGAS). Its proper motion is $\mu_\alpha \cos{\delta}\,=\,28.5\,\pm\,1.9$~mas~yr$^{-1}$ and $\mu_\delta\,=\,11.97\,\pm\,0.76$~mas~yr$^{-1}$. The photometric metallicity is [Fe/H]$\,=\,-0.38\,\pm\,0.5$~dex \citep{2006ApJ...638.1004A}. We classify the companion as M8, implying a spectrophotometric distance of $148\,\pm\,29$~pc ($2.5\,\sigma$ agreement with the primary distance). We measure a proper motion for the M8 of $\mu_\alpha \cos{\delta}\,=\,35\,\pm\,33$~mas~yr$^{-1}$ and $\mu_\delta\,=\,93\,\pm\,32$~mas~yr$^{-1}$. As a result, the false alarm probability is $2\,\times\,10^{-4}$.

	\subsection{Uncertain common proper motion systems}
		\label{uncertain_pairs}

		\subsubsection{TYC~54-833-1 + ULAS~J02553253+0532122}
			The primary is an F5V at a distance of $252^{+19}_{-16}$~pc \citep{2016A&A...595A...4L}. Its proper motion is $\mu_\alpha \cos{\delta}\,=\,28.0\,\pm\,1.5$~mas~yr$^{-1}$ and $\mu_\delta\,=\,29.8\,\pm\,0.8$~mas~yr$^{-1}$. We got its metallicity from \citet{2006ApJ...638.1004A}, [Fe/H]$\,=\,0.06\,\pm\,0.21$~dex. The companion is a L0, observed with LIRIS (see Figure~\ref{wht_spectra}), at a spectrophotometric distance of $139\,\pm\,26$~pc. This object is undetected in 2MASS so its proper motion is based on its SDSS and ULAS position only, resulting in a shorter base line and larger uncertainties. The proper motion we measured is $\mu_\alpha \cos{\delta}\,=\,28\,\pm\,30$~mas~yr$^{-1}$ and $\mu_\delta\,=\,40\,\pm\,30$~mas~yr$^{-1}$. Due to the large uncertainties on the proper motion of the L dwarf the false alarm probability for this pair is $1\,\times\,10^{-4}$.

		\subsubsection{TYC~5163-1356-1 + SDSS~J202410.30-010039.2}
			The primary is an F2 from the Tycho catalogue, with an estimated photometric metallicity of $0.47\,\pm\,0.30$~dex \citep{2006ApJ...638.1004A}. Its distance and proper motions (from TGAS) are d$\,=\,198^{+13}_{-11}$~pc, $\mu_\alpha \cos{\delta}\,=\,-11.49\,\pm\,0.77$~mas~yr$^{-1}$ and $\mu_\delta\,=\,-18.70\,\pm\,0.67$~mas~yr$^{-1}$. The companion is an M9 dwarf, selected by cross matching 2MASS and SDSS. Because of the relatively short baseline between the two available epochs and the large centroiding uncertainty for the 2MASS epoch (our target is at the limit for detection), the resulting proper motion has large uncertainties, $\mu_\alpha \cos{\delta}\,=\,-19\,\pm\,46$~mas~yr$^{-1}$ and $\mu_\delta\,=\,-79\,\pm\,46$~mas~yr$^{-1}$. The spectrophotometric distance is $136\,\pm\,28$~pc, in good agreement with the parallax of the primary, and thanks to the relatively small separation of the pair ($83''$, corresponding to $\sim\,17,000$~au), the false alarm probability for this pair is $1.8\,\times\,10^{-5}$.

		\subsubsection{2MASS~J23181098+1503259 + ULAS~J23180626+1506100}
			The primary is a metal-rich late-G dwarf from the LAMOST~DR2 catalogue ([Fe/H]$\,=\,0.36\,\pm\,0.10$~dex). Its spectrophotometric distance places it at $217\,\pm\,15$~pc. This star is too faint to be in TGAS \citep[$V\,=\,12.987$~mag,][]{2016yCat.2336....0H}, so we got its proper motion from UCAC4 \citep{2013AJ....145...44Z}. The two components are $\mu_\alpha \cos{\delta}\,=\,9.4\,\pm\,1.7$~mas~yr$^{-1}$ and $\mu_\delta\,=\,-7.1\,\pm\,2.2$~mas~yr$^{-1}$. The companion is an M9, at a spectrophotometric distance of $216\,\pm\,41$~pc, and with a very uncertain proper motion of $\mu_\alpha \cos{\delta}\,=\,47\,\pm\,24$~mas~yr$^{-1}$ and $\mu_\delta\,=\,38\,\pm\,24$~mas~yr$^{-1}$. The false alarm probability for this pair is $1.3\,\times\,10^{-4}$.

		\onecolumn
		\begin{deluxetable}{l c c c c c c c c c}
 			
            \tablecolumns{10}
            \tablewidth{0pt}
            \tabletypesize{\scriptsize}

            \tablehead{
 				\colhead{ID} & \colhead{R.A.} & \colhead{Dec.} & \colhead{SpT} & \colhead{$J$} & \colhead{$s$} & \colhead{CGB?} & \colhead{Consistent} & \colhead{Consistent} & \colhead{Binary?}  \\
 				\colhead{} & \colhead{[hh:mm:ss.ss]} & \colhead{[dd:mm:ss.s]} & \colhead{} & \colhead{[mag]} & \colhead{[kau]} & \colhead{} & \colhead{distance?} & \colhead{PM?} & \colhead{}
            }

 			\tablecaption{The new candidate systems assessed here. \label{newsystems}}

            \startdata
 			ULAS~J0008+0806 & 00:08:12.84 & +08:06:42.1 & M9 & 16.560$\pm$0.017 & \multirow{2}{*}{\ldots} & \multirow{2}{*}{No} & \multirow{2}{*}{No} & \multirow{2}{*}{No} & \multirow{2}{*}{N} \\
 			BD+07~3	& 00:08:05.09 & +08:05:59.2 & F2V  & \textit{9.322$\pm$0.024} & & & & & \\
 			\hline
 			ULAS~J0015+0248 & 00:15:14.83 & +02:48:03.9 & L1 & 15.757$\pm$0.007 & \multirow{2}{*}{1.8} & \multirow{2}{*}{Yes} & \multirow{2}{*}{Yes} & \multirow{2}{*}{Yes} & \multirow{2}{*}{N} \\
 			2MASS~J0015+0247 & 00:15:15.62 & +02:47:37.4 & K7V & \textit{9.524$\pm$0.035} & & & & & \\
 			\hline
 			ULAS~J0122+0705 & 01:22:37.06 & +07:05:57.9 & L0 & 17.092$\pm$0.021 & \multirow{2}{*}{\ldots} & \multirow{2}{*}{No} & \multirow{2}{*}{No} & \multirow{2}{*}{Yes} & \multirow{2}{*}{N} \\
 			TYC~27-721-1 & 01:22:29.75 & +07:06:13.8 & K1V & \textit{8.245$\pm$0.030} & & & & & \\
 			\hline
 			ULAS~J0255+0532 & 02:55:32.53 & +05:32:12.2 & L0 & 17.240$\pm$0.026 & \multirow{2}{*}{29} & \multirow{2}{*}{No} & \multirow{2}{*}{Yes} & \multirow{2}{*}{Yes} & \multirow{2}{*}{U} \\
 			TYC~54-833-1 & 02:55:38.23 & +05:29:46.2 & F5V & \textit{8.849$\pm$0.026} & & & & & \\
 			\hline
 			ULAS~J0324+0457 & 03:24:41.33 & +04:57:52.0 & M9 & 16.780$\pm$0.016 & \multirow{2}{*}{\ldots} & \multirow{2}{*}{No} & \multirow{2}{*}{No} & \multirow{2}{*}{Yes} & \multirow{2}{*}{N} \\
 			BD+04~533 & 03:24:46.46 & +04:55:17.8 & G5V & \textit{7.741$\pm$0.026} & & & & & \\
 			\hline
 			ULAS~J0741+2316 & 07:41:04.39 & +23:16:37.6 & L0 & 16.036$\pm$0.006 & \multirow{2}{*}{\ldots} & \multirow{2}{*}{No} & \multirow{2}{*}{No} & \multirow{2}{*}{No} & \multirow{2}{*}{N} \\
 			TYC~1912-724-1	& 07:41:11.84 & +23:15:36.1 & F5III & \textit{10.855$\pm$0.022} & & & & & \\
 			\hline
 			ULAS~J0744+2513 & 07:44:36.00 & +25:13:30.6 & M8 & 16.676$\pm$0.011 & \multirow{2}{*}{\ldots} & \multirow{2}{*}{No} & \multirow{2}{*}{No} & \multirow{2}{*}{No} & \multirow{2}{*}{N} \\
 			TYC~1916-1611-1 & 07:44:37.79 & +25:13:41.1 & F6V & \textit{9.341$\pm$0.024} & & & & & \\ 
 			\hline
 			ULAS~J0836+0221 & 08:36:13.47 & +02:21:06.3 & M8 & 14.707$\pm$0.010 & \multirow{2}{*}{14} & \multirow{2}{*}{No} & \multirow{2}{*}{Yes} & \multirow{2}{*}{No} & \multirow{2}{*}{N} \\
 			BD+02~2020 & 08:36:12.01 & +02:21:24.6 & K0V & \textit{8.891$\pm$0.029} & & & & & \\ 
 			\hline
 			ULAS~J0900+2930 & 09:00:04.74 & +29:30:22.1 & L0 & 17.993$\pm$0.032 & \multirow{2}{*}{16} & \multirow{2}{*}{Yes} & \multirow{2}{*}{Yes} & \multirow{2}{*}{Yes} & \multirow{2}{*}{R} \\
 			NJCM~J0900+2931 & 09:00:13.50 & +29:31:20.3 & M3.5 & 13.638$\pm$0.002 & & & & & \\ 
 			\hline
 			ULAS~J0936+3115 & 09:36:13.16 & +31:15:13.5 & M7 & 16.593$\pm$0.008 & \multirow{2}{*}{5.4} & \multirow{2}{*}{No} & \multirow{2}{*}{Yes} & \multirow{2}{*}{No} & \multirow{2}{*}{N} \\
 			NJCM~J0936+3116 & 09:36:16.58 & +31:16:36.8 & M3.5 & \textit{11.742$\pm$0.021} & & & & & \\ 
 			\hline
 			ULAS~J0938+0815 & 09:38:36.78 & +08:15:11.0 & M7 & 14.366$\pm$0.002 & \multirow{2}{*}{\ldots} & \multirow{2}{*}{No} & \multirow{2}{*}{No} & \multirow{2}{*}{No} & \multirow{2}{*}{N} \\
 			TYC~821-1173-1 & 09:38:41.46 & +08:14:56.7 & F6V & \textit{10.605$\pm$0.022} & & & & & \\
 			\hline
 			ULAS~J0949-0015 & 09:49:36.26 & -00:15:28.8 & M9 & 15.710$\pm$0.008 & \multirow{2}{*}{\ldots} & \multirow{2}{*}{No} & \multirow{2}{*}{No} & \multirow{2}{*}{No} & \multirow{2}{*}{N} \\
 			IDS~09445+0011~AB & 09:49:38.13 & -00:16:32.7 & G0V & \textit{8.376$\pm$0.021} & & & & & \\
 			\hline
 			ULAS~J1003+0511 & 10:03:37.92 & +05:11:41.7 & M8 & 18.236$\pm$0.053 & \multirow{2}{*}{35} & \multirow{2}{*}{No} & \multirow{2}{*}{Yes} & \multirow{2}{*}{No} & \multirow{2}{*}{N} \\
 			BD+05~2275 & 10:03:38.19 & +05:14:16.9 & F5V & \textit{9.723$\pm$0.023} & & & & & \\
 			\hline
 			ULAS~J1102+1040 & 11:02:51.03 & +10:40:46.6 & M8 & 16.507$\pm$0.015 & \multirow{2}{*}{6.1} & \multirow{2}{*}{No} & \multirow{2}{*}{Yes} & \multirow{2}{*}{No} & \multirow{2}{*}{N} \\ 
 			2MASS~J1102+1041 & 11:02:55.20 & +10:41:03.6 & K7V & \textit{10.518$\pm$0.022} & & & & & \\
 			\hline
 			ULAS~J1217+1427 & 12:17:36.43 & +14:27:11.7 & L1 & 15.995$\pm$0.008 & \multirow{2}{*}{2.7} & \multirow{2}{*}{Yes} & \multirow{2}{*}{Yes} & \multirow{2}{*}{Yes} & \multirow{2}{*}{R} \\
 			HD~106888 & 12:17:36.25 & +14:26:34.5 & F8V & \textit{7.125$\pm$0.020} & & & & & \\
 			\hline
 			ULAS~J1219+0154 & 12:19:32.54 & +01:54:33.0 & M9 & 14.977$\pm$0.004 & \multirow{2}{*}{0.7} & \multirow{2}{*}{Yes} & \multirow{2}{*}{Yes} & \multirow{2}{*}{Yes} & \multirow{2}{*}{R} \\
 			PYC~12195+0154 & 12:19:33.16 & +01:54:26.8 & M3V & \textit{10.543$\pm$0.022} & & & & & \\
 			\hline
 			ULAS~J1222+1407 & 12:22:59.30 & +14:07:50.1 & L0 & 17.979$\pm$0.053 & \multirow{2}{*}{6.7} & \multirow{2}{*}{Yes} & \multirow{2}{*}{Yes} & \multirow{2}{*}{Yes} & \multirow{2}{*}{R} \\ 
 			NJCM J1222+1407	& 12:22:57.28 & +14:07:18.5 & M4V & 14.431$\pm$0.003 & & & & & \\
 			\hline
 			ULAS~J1224+2453 & 12:24:16.99 & +24:53:33.4 & M7 & 15.320$\pm$0.010 & \multirow{2}{*}{\ldots} & \multirow{2}{*}{No} & \multirow{2}{*}{No} & \multirow{2}{*}{No} & \multirow{2}{*}{N} \\
 			TYC~1989-265-1 & 12:24:21.41 & +24:52:53.9 & F5V & \textit{10.210$\pm$0.028} & & & & & \\
 			\hline
 			ULAS~J1330+0914 & 13:30:02.49 & +09:14:32.1 & L2 & 17.983$\pm$0.026 & \multirow{2}{*}{61} & \multirow{2}{*}{Yes} & \multirow{2}{*}{Yes} & \multirow{2}{*}{Yes} & \multirow{2}{*}{R} \\
 			TYC~892-36-1 & 13:29:58.74 & +09:10:17.7 & G5V & \textit{10.918$\pm$0.023} & & & & & \\
 			\hline
 			ULAS~J1342+2933 & 13:42:01.99 & +29:33:40.0 & M8 & 15.966$\pm$0.008 & \multirow{2}{*}{\ldots} & \multirow{2}{*}{No} & \multirow{2}{*}{No} & \multirow{2}{*}{No} & \multirow{2}{*}{N} \\
 			BD+30~2436 & 13:41:33.04 & +29:39:31.5 & K0V & \textit{7.872$\pm$0.021} & & & & & \\
 			\hline
 			ULAS~J1345-0058 & 13:45:12.42 & -00:58:44.3 & M8 & 16.210$\pm$0.008 & \multirow{2}{*}{9.1} & \multirow{2}{*}{No} & \multirow{2}{*}{Yes} & \multirow{2}{*}{No} & \multirow{2}{*}{N} \\ 
 			NJCM J1345-0057	& 13:45:18.73 & -00:57:29.5 & M5V & 13.956$\pm$0.002 & & & & & \\
 			\hline
 			ULAS~J1500+1302 & 15:00:10.74 & +13:02:12.2 & M8 & 16.755$\pm$0.010 & \multirow{2}{*}{\ldots} & \multirow{2}{*}{No} & \multirow{2}{*}{No} & \multirow{2}{*}{No} & \multirow{2}{*}{N} \\
 			HD~132681 & 15:00:07.46 & +13:01:48.1 & G5V & \textit{8.536$\pm$0.029} & & & & & \\
 			\hline
 			ULAS~J1522-0136 & 15:22:46.58 & -01:36:42.6 & M9.5 & 18.189$\pm$0.064 & \multirow{2}{*}{57} & \multirow{2}{*}{Yes} & \multirow{2}{*}{Yes} & \multirow{2}{*}{Yes} & \multirow{2}{*}{R} \\
 			HIP~75262 & 15:22:40.18 & -01:32:29.6 & F5V & \textit{8.539$\pm$0.019} & & & & & \\
 			\hline
 			ULAS~J1540+0102 & 15:40:05.10 & +01:02:08.8 & M9 & 15.127$\pm$0.005 & \multirow{2}{*}{0.6} & \multirow{2}{*}{Yes} & \multirow{2}{*}{Yes} & \multirow{2}{*}{Yes} & \multirow{2}{*}{R} \\
 			NJCM J1540+0102 & 15:40:05.91 & +01:02:15.1 & M3V & 11.696$\pm$0.001 & & & & & \\
 			\hline
 			ULAS~J1600+2843 & 16:00:36.55 & +28:43:06.2 & L8 & 17.651$\pm$0.030 & \multirow{2}{*}{\ldots} & \multirow{2}{*}{No} & \multirow{2}{*}{No} & \multirow{2}{*}{No} & \multirow{2}{*}{N} \\
 			TYC~2041-1324-1	& 16:00:33.65 & +28:44:27.5 & F6V & \textit{9.695$\pm$0.023} & & & & & \\
 			\hline
 			ULAS~J1606+2634 & 16:06:11.53 & +26:34:51.8 & M8 & 15.854$\pm$0.005 & \multirow{2}{*}{5.9} & \multirow{2}{*}{Yes} & \multirow{2}{*}{Yes} & \multirow{2}{*}{Yes} & \multirow{2}{*}{R} \\
 			TYC~2038-524-1 & 16:06:10.36 & +26:34:03.6 & G8V & \textit{8.729$\pm$0.027} & & & & & \\
 			\hline
 			SDSS~J1724+2336 & 17:24:37.52 & +23:36:49.3 & M7 & \textit{15.678$\pm$0.062} & \multirow{2}{*}{\ldots} & \multirow{2}{*}{No} & \multirow{2}{*}{No} & \multirow{2}{*}{No} & \multirow{2}{*}{N} \\
 			TYC~2074-442-1 & 17:24:35.20 & +23:34:58.8 & F6V & \textit{8.901$\pm$0.027} & & & & & \\
 			\hline
 			SDSS~J2024-0100 & 20:24:10.30 & -01:00:39.2 & M9 & \textit{16.98$\pm$0.18} & \multirow{2}{*}{16} & \multirow{2}{*}{No} & \multirow{2}{*}{Yes} & \multirow{2}{*}{Yes} & \multirow{2}{*}{U} \\
 			BD-01~3972 & 20:24:11.40 & -01:02:00.9 & F2V & \textit{8.616$\pm$0.029} & & & & & \\
 			\hline
 			ULAS~J2318+1506 & 23:18:06.26 & +15:06:10.0 & M9 & 17.983$\pm$0.050 & \multirow{2}{*}{36} & \multirow{2}{*}{No} & \multirow{2}{*}{Yes} & \multirow{2}{*}{Yes} & \multirow{2}{*}{U} \\
 			2MASS~J2318+1503 & 23:18:11.00 & +15:03:26.0 & G8V & 11.482$\pm$0.001 & & & & & \\
 			\hline
 			ULAS~J2338+1216 & 23:38:39.81 & +12:16:34.1 & M8 & 16.915$\pm$0.017 & \multirow{2}{*}{76} & \multirow{2}{*}{Yes} & \multirow{2}{*}{Yes} & \multirow{2}{*}{Yes} & \multirow{2}{*}{R} \\
 			TYC~1172-357-1 & 23:38:30.04 & +12:15:22.9 & F6V & \textit{10.486$\pm$0.024} & & & & &
            \enddata

            \tablecomments{$J$ band magnitudes are from the UKIDSS LAS, except values in italics which are from 2MASS. For the common distance pairs, $s$ is calculated using the average distance to the pair. The last four columns summarize the outcome of the companionship assessment described in Section~\ref{new_systems}. ``R'' stands for ``robust system'', ``U'' for ``uncertain system'', and ``N'' for ``not a binary''.}
		\end{deluxetable}
		\twocolumn

\section{Discussion and future work}
	\label{discussion}
	We now broadly consider our simulation and initial follow-up sample together, in the context of CGBs subsample completeness/uniformity and observational follow-up economy.

	Our previous discussion has shown that the parameter space of the full CGB population covers a broad range of UCD properties: $0.02 \le M/M_{\odot} \le 0.1$, $0.1 \le {\rm age/Gyr} \le 14$, $400 \le T_{\rm eff}/{\rm K} \le 2800$ and $-2.5 \le {\rm [Fe/H]} \le 0.5$. And that effectively exploring this parameter space requires targeting certain subsets (see Table~\ref{sim_res_summary}). However, the number of CGBs in these subsets is non-uniform, and an economic follow-up plan should pursue only fractions of the larger subsets. One can aim for a roughly equal split between young/old and metal-rich/metal-poor systems by targeting: 
	(i) all the metal-rich L/T CGBs and $\sim$10\% of the metal-rich M CGBs, (ii) all the ancient metal-poor L/T CGBs in the thick disk/halo, and a few \% of the metal poor M CGBs, (iii) all the young CGBs, and (iv) all/any Y dwarf CGBs. This collectively represents a more uniform sample of $\sim$500 \emph{Gaia} benchmark systems (including $\sim$200 late M CGBs, close to 300 L type CGBs, 10-15 T types, and a few Y type CGBs).

	The full observational search-and-follow-up campaign for CGBs can be an expensive task, in particular when targeting faint UCDs. We discussed in Section~\ref{priority} how to improve the efficiency of such a campaign, and we outlined the following set of guidelines, that we hope can be useful to design future efforts in this direction.
	
	\begin{itemize}
  		\item{In general, near-infrared surveys such as UKIDSS, UHS, and VISTA can yield almost all CGBs.}
  		\item{The AllWISE database adds a relatively small number of L type CGBs beyond what can be identified using near-infrared surveys.}
  		\item{The combination of AllWISE and NEOWISE imaging could improve mid-IR sensitivity and yield a sample of $\sim\,1-3$ Y-dwarf CGBs.}
  		\item{Most CGBs can be identified out to an angular separation of $3-5$~arcmin for M and L dwarfs, and $\sim\,$15~arcmin for T dwarfs.}
  		\item{Ultracool halo M dwarf CGBs all have angular separations $\rho\,<\,1$~arcmin.}
  		\item{Priority subsets can be sought in different ways. One could search for CGB candidates around target primaries (i.e. sub-giants, white dwarfs, metal rich/poor main sequence stars, young stars, thick disk and halo stars).}
  		\item{Comprehensive lists of interesting primaries may come from various sources, including \emph{Gaia} photometric and spectroscopic analysis.}
  		\item{As a complementary approach (assuming non-comprehensive primary information), one could seek observationally unusual UCDs (e.g. colour or spectroscopic outliers) that may be more likely members of our priority subsets, and search around them for primary stars.}
  		\item{A follow-up programme to confirm CGBs can be guided by Figure~\ref{dist_pm_sep}. Candidate CGBs will have known angular separation from their potential primaries, and \emph{Gaia} will provide distance constraints and total proper motions for these primaries. CGB confirmation is likely to require only common-distance (i.e. by measuring UCD spectral type) if $\rho\,<\,6$~arcsec, and in most cases when $d\,<\,30$~pc; beyond these limits common-proper-motion will be needed, provided that the proper motion is significant ($\mu\,\geq\,20\,-\,25$~mas~yr$^{-1}$); for low proper motion systems ($\mu\,<\,20\,-\,25$~mas~yr$^{-1}$) common-RV may be required.}
	\end{itemize}
	
	The benchmark search reported here represents only a small portion of the full \emph{Gaia} benchmark population, but it has a broadly similar character to the bulk of our simulation (i.e. the late M and L dwarfs): $J<18$~mag (simulation cf. $J<19$~mag), proper motion $>$14~mas~yr$^{-1}$ (cf. $>$5~mas~yr$^{-1}$), angular separation $\le\,3$~arcmin (cf. $\le\,5$~arcmin), distance $<\,230$~pc (cf. L dwarfs $<\,300$~pc; M dwarfs $<\,550$~pc), projected separation $<\,$76~kau (cf. $<\,$50~kau), [Fe/H] between $-0.39$ and $+0.36$~dex (cf. between $-1.0$ and $+0.4$~dex). And our comprehensive follow-up of the initial candidate sample thus means the results can usefully inform larger searches going forward.

	Our initial candidate companions had a low rate of (non-UCD) contamination, with 94\% spectroscopically confirmed as UCDs. But 40\% of the spectroscopically confirmed UCDs did not pass our common distance test, i.e. a significant fraction of false positive chance alignments. Access to released \emph{Gaia} parallaxes will improve this somewhat, but the UCD photometric distance uncertainties are the most significant factor leading to chance alignments. Of the pairs that pass our common-distance test, the majority (65\%) pass the common-proper-motion test. And of those that fail, the UCDs have much higher proper motion than their candidate primaries in roughly half the cases, with similar proper motion magnitude but different direction accounting for the remaining cases. We also note that there are just 2/10 pairs that pass the common-proper-motion test that did not pass the common-distance test.

	In the future, follow-up could be economized through improved photometric spectral types and distance constraints for the candidate companions \citep[e.g.][]{2016A&A...589A..49S}, though care must be taken to allow for UCDs with non-standard properties/colours (that may affect the distance estimates). Early proper motion measurements of the UCD candidates could also improve follow-up economy, since most rejected UCD companion candidates show significant motion. We also note that four of our final selected systems actually gained CGB statistical status without the need for common-proper motion assessment. These were the systems with the smallest projected separations ($\le\,$2.7 kau), and proper motion measurements for such UCDs could be de-prioritised. None of our newly presented systems required common-radial velocity confirmation, and our simulation predicts that only $\sim3\%$ of CGBs should require such follow-up. This is important to note for future search-and-follow-up campaigns, since radial velocity is extremely time consuming to obtain for such faint UCDs. Overall, our approach of filtering candidates, via liberal common-distance and common-proper-motion cuts prior to statistical tests, seems to be fit-for-purpose. While the ten robust systems have very low false alarm probabilities, three less certain systems passed through our filters but were then flagged as statistically marginal. Thus our approach should not be missing genuine benchmarks, and effectively identified marginal cases that could be further analysed through additional follow-up (i.e. improved proper motion measurements and/or radial velocities of faint companions, and refined constraints on the metallicity of the primaries).

	We have yet to fully explore age constraints for our initial benchmark sample, but broad Bayesian analysis for individual primaries will accompany future \emph{Gaia} releases \citep[e.g.][]{2013A&A...559A..74B}. In addition, dynamical mass constraints may be available for some UCD companions ($\sim$30\% of these UCD components are themselves expected to be unresolved binaries) through adaptive optics and/or radial velocity follow-up.

	A full CGB population has huge potential to improve the accuracy with which we can calibrate the properties of ultracool atmospheres and low-mass objects: (i) it would offer extensive ``bench-tests'' for theoretical models \citep[e.g.][]{2012ApJ...756..172M,2014IAUS..299..271A,2016ApJ...817L..19T} and retrieval techniques \citep[e.g.][]{2017arXiv170101257B}, (ii) could guide classification of the lowest mass objects \citep[e.g.][]{2007ApJ...669.1235L,2017MNRAS.464.3040Z}, (iii) aid the identification of interesting/unusual free-floating UCDs, (iv) inform population synthesis helping to constrain the initial mass function and formation history of Galactic brown dwarfs \citep[e.g.][]{2013MNRAS.430.1171D,2015MNRAS.449.3651M}, and (v) help interpret the properties of giant planets \citep[e.g.][]{2010SPIE.7735E..2EK}.

\section*{Acknowledgments} 
	We thank the anonymous referee for comments that have significantly improved the quality of this manuscript. FM, DJP, NJC, and HRAJ acknowledge support from the UK's Science and Technology Facilities Council, grant numbers ST/M001008/1, ST/N001818/1, and ST/J001333/1. Partly based on observations made with the Gran Telescopio Canarias (GTC), installed at the Spanish Observatorio del Roque de los Muchachos of the Instituto de Astrofísica de Canarias, in the island of La Palma. Partly based on observations made with the 4.2~m William Herschel Telescope operated on the island of La Palma by the ING at the Observatorio del Roque de los Muchachos of the Instituto de Astrof\`isica de Canarias. This research has benefitted from the SpeX Prism Spectral Libraries, maintained by Adam Burgasser at \url{http://pono.ucsd.edu/~adam/browndwarfs/spexprism}. This publication makes use of data products from the Wide-field Infrared Survey Explorer, which is a joint project of the University of California, Los Angeles, and the Jet Propulsion Laboratory/California Institute of Technology, funded by the National Aeronautics and Space Administration. This publication makes use of data products from the Two Micron All Sky Survey, which is a joint project of the University of Massachusetts and the Infrared Processing and Analysis Center/California Institute of Technology, funded by the National Aeronautics and Space Administration and the National Science Foundation. Guoshoujing Telescope (the Large Sky Area Multi-Object Fiber Spectroscopic Telescope LAMOST) is a National Major Scientific Project built by the Chinese Academy of Sciences. Funding for the project has been provided by the National Development and Reform Commission. LAMOST is operated and managed by the National Astronomical Observatories, Chinese Academy of Sciences. Funding for SDSS-III has been provided by the Alfred P. Sloan Foundation, the Participating Institutions, the National Science Foundation, and the U.S. Department of Energy Office of Science. The SDSS-III web site is \url{http://www.sdss3.org/}. SDSS-III is managed by the Astrophysical Research Consortium for the Participating Institutions of the SDSS-III Collaboration including the University of Arizona, the Brazilian Participation Group, Brookhaven National Laboratory, Carnegie Mellon University, University of Florida, the French Participation Group, the German Participation Group, Harvard University, the Instituto de Astrofisica de Canarias, the Michigan State/Notre Dame/JINA Participation Group, Johns Hopkins University, Lawrence Berkeley National Laboratory, Max Planck Institute for Astrophysics, Max Planck Institute for Extraterrestrial Physics, New Mexico State University, New York University, Ohio State University, Pennsylvania State University, University of Portsmouth, Princeton University, the Spanish Participation Group, University of Tokyo, University of Utah, Vanderbilt University, University of Virginia, University of Washington, and Yale University.

\appendix

\section{Observations Log}
	\label{obs_log}
	
	\begin{table*}
	\begin{tabular}{l c c c c}
	Object & Instrument & Obs. Date & Exp. Time & Standard \\
	ID     &            & (YYYY-MM-DD) & (s)       & \\
	\hline
	\hline
	ULAS~J00081284+0806421 & LIRIS & 2015-11-21 & 6$\times$400 & HIP~113889 \\
	ULAS~J00151479+0248020 & LIRIS & 2015-11-21 & 4$\times$400 & HIP~117927 \\
	ULAS~J01223706+0705579 & LIRIS & 2015-11-21 & 6$\times$400 & HIP~110573 \\
	ULAS~J02553253+0532122 & LIRIS & 2015-11-21 & 6$\times$400 & HIP~10732 \\
	ULAS~J03244133+0457520 & LIRIS & 2015-11-21 & 6$\times$400 & HIP~20884 \\
	ULAS~J07410439+2316376 & LIRIS & 2015-11-21 & 6$\times$400 & HIP~33300 \\
	ULAS~J07443600+2513306 & LIRIS & 2015-11-21 & 6$\times$400 & HIP~33300 \\
	ULAS~J08361347+0221063 & LIRIS  & 2016-04-25 & 4$\times$300 & HIP~45167 \\
	ULAS~J09000474+2930221 & OSIRIS & 2015-11-30 & 900 & G~191--B2B \\
	ULAS~J09361316+3115135 & LIRIS  & 2016-04-25 & 4$\times$300 & HIP~50303 \\
	ULAS~J09383678+0815110 & LIRIS  & 2016-04-25 & 12$\times$300 & HIP~45167 \\
	ULAS~J09493641--0015334 & LIRIS & 2015-11-22 & 4$\times$400 & HIP~110573 \\
	ULAS~J10033792+0511417 & OSIRIS & 2015-11-30 & 900 & G~191--B2B \\
	ULAS~J11025103+1040466 & LIRIS  & 2016-04-24 & 12$\times$300 & HIP~52911 \\
	ULAS~J12173673+1427096 & LIRIS & 2015-11-21 & 6$\times$400 & HIP~33300 \\
	ULAS~J12193254+0154330 & LIRIS  & 2016-04-24 & 4$\times$300 & HIP~61631 \\
	PYC~12195+0154 & LIRIS  & 2016-04-24 & 4$\times$10 & HIP~52911 \\
	ULAS~J12225930+1407501 & OSIRIS & 2016-01-19 & 900 & G~191--B2B \\
	ULAS~J12241699+2453334 & LIRIS  & 2016-04-24 & 8$\times$300 & HR~4705 \\
	ULAS~J13300249+0914321 & OSIRIS & 2016-03-28 & 900 & Ross~640 \\
	ULAS~J13420199+2933400 & LIRIS  & 2016-04-24 & 12$\times$300 & HIP~12178 \\
	ULAS~J13451242--0058443 & LIRIS  & 2016-04-25 & 12$\times$300 & HIP~68498 \\
	ULAS~J15001074+1302122 & LIRIS  & 2016-04-25 & 12$\times$300 & HIP~75230 \\
	ULAS~J15224658--0136426 & OSIRIS & 2016-03-29 & 900 & Ross~640 \\
	ULAS~J15400510+0102088 & LIRIS  & 2016-04-24 & 8$\times$300 & HIP~77516 \\
	NJCM~J15400591+0102151 & LIRIS  & 2016-04-24 & 4$\times$90 & HIP~77516 \\
	ULAS~J16003655+2843062 & OSIRIS & 2016-03-28 & 900 & Ross~640 \\
	ULAS~J16061153+2634518 & LIRIS  & 2016-04-25 & 10$\times$300 & HIP~79332 \\
	SDSS~J172437.52+233649.3 & OSIRIS & 2015-09-01 & 900 & Ross~640 \\
	SDSS~J20240.30--010039.2 & OSIRIS & 2015-09-01 & 900 & Ross~640 \\
	ULAS~J23180626+1506100 & OSIRIS & 2015-09-01 & 900 & Ross~640 \\
	ULAS~J23383981+1216341 & LIRIS & 2015-11-21 & 4$\times$400 & HIP~110573 \\
	\hline
	\hline
	\end{tabular}
	\caption{The observing log for the spectra presented here. For each object we present the instrument used, the date of observation, the exposure time, and the standard star used for flux calibration and telluric correction. \label{obs_log_table}}
	\end{table*}

\label{lastpage}

\end{document}